\documentclass[journal=nalefd,manuscript=article]{achemso}

\usepackage[utf8]{inputenc}
\usepackage[version=3]{mhchem} 
\usepackage{xcolor}
\usepackage{graphicx}
\usepackage{dcolumn}
\usepackage{bm}
\usepackage{braket}
\usepackage{mathbbol}
\usepackage{gensymb}

\author{Marco Romanelli}
\affiliation{Department of Chemical Sciences, University of Padova, via Marzolo 1, 35131 Padova, Italy}
\author{Gabriel Gil}
\affiliation{Dipartimento di Scienze e Innovazione Tecnologica, Università degli Studi del Piemonte Orientale Amedeo Avogadro, Alessandria, Italy}
\alsoaffiliation{Department of Chemical Sciences, University of Padova, via Marzolo 1, 35131 Padova, Italy}
\author{Stefano Corni}
\email{stefano.corni@unipd.it}
\affiliation{Department of Chemical Sciences, University of Padova, via Marzolo 1, 35131 Padova, Italy}
\alsoaffiliation{CNR Institute of Nanoscience, via Campi 213/A, 41125 Modena, Italy}
\alsoaffiliation{Padua Quantum Technologies Research Center, University of Padova, 35131 Padova, Italy}

\title{Quantized plasmon modes for metallic nanoparticles of arbitrary shape with a generic dielectric function}


\begin{document}

\begin{tocentry}
\centering
    \includegraphics{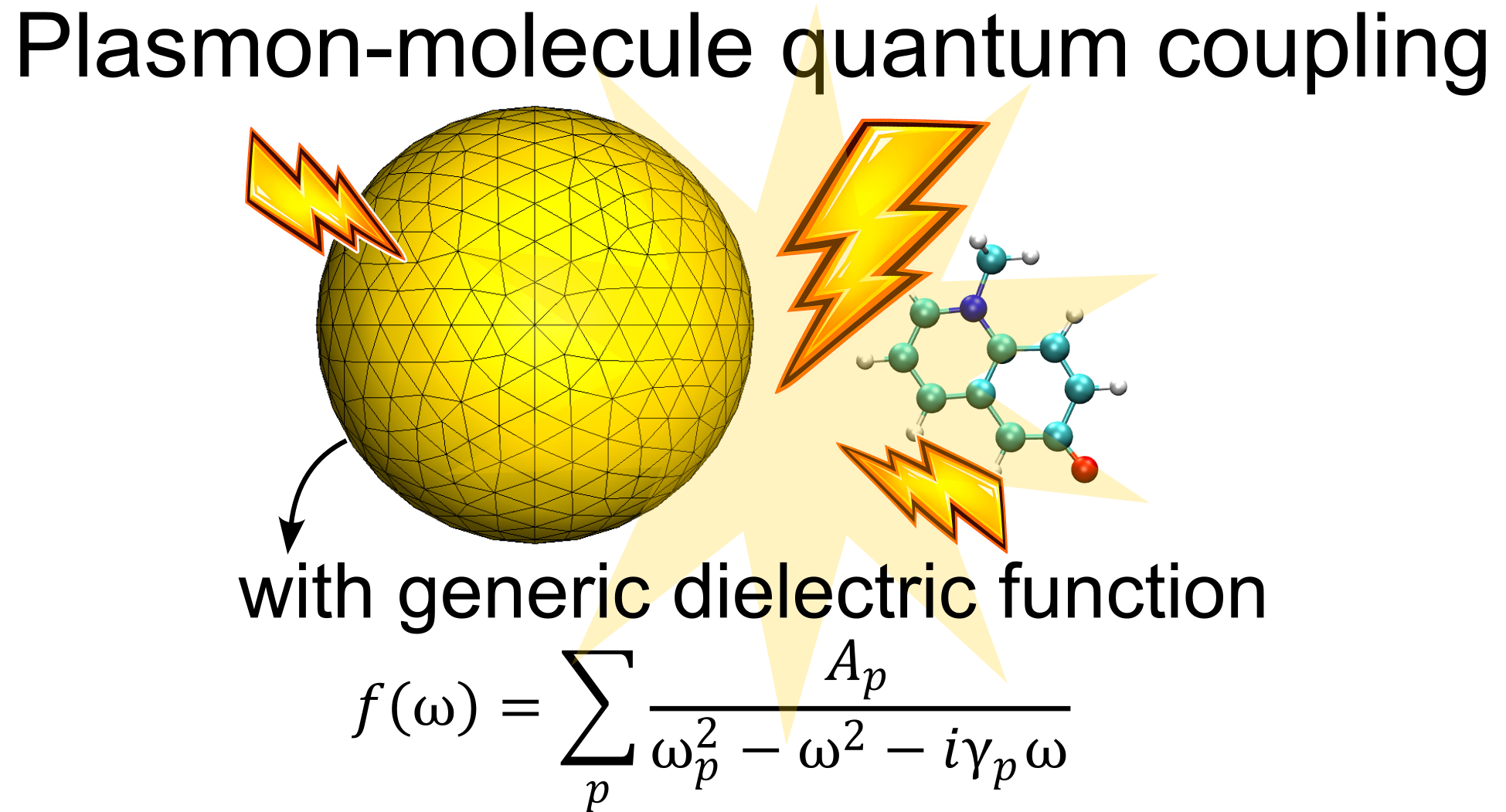}
\end{tocentry}

\begin{abstract}
In this work we introduce an effective approach to quantize the electromagnetic response of plasmonic metallic nanostructures. 
Their shape is arbitrary and they feature a realistic description of the frequency-dependent metal dielectric function that is based on experimental data. 
The derived quantum modes correctly reproduce the linear response macroscopic polarization of the nanoparticle upon external drive according to classical macroscopic Maxwell equations in the quasistatic limit.
We further investigate the coupling of these modes to a quantum-chemical molecular description.
The presented methodology paves the way for accurate modeling of plexcitonic system, where strong plasmon-molecule coupling and/or strong-driving fields call for a quantized description of the plasmonic response. 
\end{abstract}

Plasmonic metallic nanoparticles (NPs) exhibit coherent and collective oscillations of the metal conduction electrons upon light irradiation. 
Such phenomenon, referred to as Localized Surface Plasmon Resonance (LSPR), leads to the enhancement of the external electromagnetic field in the proximity of the metal surface, down to molecular scale\cite{giannini2011,mayer2011,murray2007}. 
Over the past decades there has been rising interest in making use of NPs to shape molecular properties by coupling electronic transitions of molecules with the plasmon-enhanced local electromagnetic field arising from LSPRs. 
This area, known as molecular plasmonics\cite{della2013handbook}, has proved to be an effective non-invasive way of modulating molecular properties as a result of light-matter coupling at the nanoscale.
Different applications have been illustrated, showcasing the capability of plasmonic systems to sizably modify molecular photoluminescence\cite{lakowicz2004,lakowicz2005,anger2006enhancement,yang2020,romanelli2021}, Raman scattering\cite{lee2019,dall2024,wang2020}, molecular energy transfer\cite{cao2021,imada2017single,kong2022,coane2024} and excited-state decay\cite{felicetti2020,torres2021,kuttruff2022}, just to name a few. 

Usually, due to the size difference between molecules and metallic NPs, state of the art methods\cite{mennucci2019,fregoni2022,capone2024} tackle these systems by means of multiscale approaches, where the NP response is described by classical electrodynamics while molecules are modeled at ab initio level. 
Such classical modeling of the NP is expected to break down under strong plasmon-molecule coupling and/or beyond the linear excitation regime, where strong driving fields are considered. 
Under these regimes, a quantized description of the NP response is unavoidable to capture the correct system dynamics\cite{waks2010,romanelli2024}.

In the following work, building on a Boundary Element Method (BEM) approach to solve classical macroscopic Maxwell equations in the quasistatic limit, we introduce effective quantum modes to model the optical response of arbitrarily-shaped NPs described by a generic dielectric function.

Similar approaches based on continuum solutions of the electrodynamics problem were previously introduced\cite{trugler2008,fuchs1975,ouyang1989,de1997} and later successfully coupled to a quantum chemistry molecular description\cite{neuman2018,fregoni2021}. 
However, in the latter case the metal dielectric function is typically described by a simple Drude or Drude–Lorentz (DL) form, thereby constituting a rough approximation to real metal dielectric functions. 
Indeed, it is well-known that for widely used plasmonic metals, like silver or gold, multiple interband transitions fall in the same spectral region of free-electrons plasmonic resonances, thus making those analytical models inaccurate to properly describe the real dielectric response. 
To overcome this limitation, the generic dielectric function approach, which relies on fitting a sum of DL oscillators to experimental values of frequency-dependent metal dielectric functions, was recently introduced in a BEM framework\cite{dall2020real}, opening up the possibility to investigate the electron dynamics of molecules in realistic plasmonic environments. 
To date, the application of this generic dielectric function approach has been restricted to deal with the classical BEM problem, and its quantized extension is the subject of the present work. 

While macroscopic quantum electrodynamics (QED) provides a way to quantize the electromagnetic field (EM) in such environments by introducing a continuum of harmonic oscillators\cite{huttner1992}, its practical application is limited to cases where perturbative approximations are valid. 
Indeed, recent works\cite{medina2021,sanchez2022,sanchez2024} have shown that such complex EM continuum structure can be effectively represented by a set of discrete modes that are lossy and coupled, and stem from a system-specific fitting procedure based on the spectral density of the photonic/plasmonic environment of interest. 
We hereby show that such few-modes effective quantization of the EM continuum via discrete, coupled and lossy modes naturally arise in a quasistatic BEM approach in a way that depends on the dielectric response of the bulk metal, with the effect of the NP shape accounted for numerically but without the need for a fitting procedure. 
This leads to the identification of the linear response matrix of a quantum system from which quantization of the NP response is achieved. 
This result shows that the possibility of writing the response of a NP in terms of a set of discrete, coupled and lossy modes is a general feature of nanostructures, at least in the quasi-static response framework.

In the quasistatic BEM approach that we consider, also termed Polarizable Continuum Model - Nanoparticle (PCM-NP)\cite{mennucci2019}, the linear response polarization of a given NP due to an external electric field is expressed in terms of a surface charge density lying on the NP surface. 
The problem of finding such surface charge density for a given driving field is solved numerically  using a BEM strategy based on a tessellation (discretization) of the NP surface. 
The BEM response equation reads,
\begin{equation}\label{eq:pcm_np}
    \bm{q}(\omega)=\bm{Q}(\omega)\bm{V}(\omega)
\end{equation}
where $\bm{q}(\omega),\bm{V}(\omega)$ are vectors collecting the polarization charges and the electrostatic potential associated to the external electric field acting at each surface element, respectively. 
On the other hand, $\bm{Q}(\omega)$ is the PCM-NP response matrix defined as 
\begin{equation}\label{eq:Q}
    \bm{Q}(\omega)=-\bm{S}^{-1}\left(2\pi\frac{\epsilon(\omega)+1}{\epsilon(\omega)-1}+\bm{DA}\right)^{-1}\left(2\pi\mathbb{I}+\bm{DA}\right)
\end{equation}
where $\bm{S}$ and $\bm{D}$ are the Calderon matrices with elements\cite{tomasi2005} 
\begin{eqnarray}\label{eq:calderon}
&& S_{ij}=\frac{1}{|\vec{s}_i-\vec{s}_j|}  \\
&& S_{ii}=1.0694\sqrt{A_i/4\pi} \\
&& D_{ij}=\frac{(\vec{s}_i-\vec{s}_j)\cdot\vec{n}_j}{|\vec{s}_i-\vec{s}_j|^3}\\
&& D_{ii}=-(2\pi+\sum_{k\neq i}D_{ik}A_k)1/A_i \, .
\end{eqnarray}
The $\bm{A}$ matrix stores the area of each surface element (also called "tessera") of the NP discretized surface, whereas $\epsilon(\omega)$ is the frequency-dependent dielectric function defining the optical dielectric response.

Following Ref.\citenum{dall2020real}, Eqs.\ref{eq:pcm_np}-\ref{eq:Q} are recast as 
\begin{equation}\label{eq:gen}
    \bm{q}(\omega)=-\frac{1}{2\pi}f(\omega)\left[\bm{A}\bm{D}^{\dagger}\bm{q}(\omega)+\bm{S}^{-1}\left(2\pi\mathbb{I}+\bm{DA}\right)\bm{V}(\omega)\right]
\end{equation}
with 
\begin{equation}\label{eq:f_gen}
f(\omega)=\frac{\epsilon(\omega)-1}{\epsilon(\omega)+1} \, .
\end{equation}
For a specific metal, the experimental $f(\omega)$ data is fitted to a sum of N DL-like poles,
\begin{equation}\label{eq:f_fit}
f(\omega)=\frac{\epsilon(\omega)-1}{\epsilon(\omega)+1}\approx\sum_p^N\frac{A_p}{\omega_p^2-\omega^2-i\gamma_p\omega}\,.
\end{equation}
By introducing the fitting functional form of $f(\omega)$ into Eq.\ref{eq:gen}, the polarization charges can be decomposed as pole-dependent charges, i.e. $\bm{q}(\omega)=\sum_p^N\bm{q}_p(\omega)$, leading to a matrix equation for each pole-dependent charge vector $\bm{q}_p(\omega)$:
\begin{eqnarray}\label{eq:gen_q}
\frac{2\pi}{A_p}(\omega_p^2-\omega^2-i\gamma_p\omega)\bm{q}_p(\omega)=-\bigr[ \bm{A}\bm{D}^{\dagger}\sum_{p'}^N\bm{q}_{p'}(\omega)+  
 \bm{S}^{-1}\left(2\pi\mathbb{I}+\bm{DA}\right)\bm{V}(\omega) \bigr] \, . 
\end{eqnarray}
As such, the PCM-NP linear response equation within the generic dielectric function approach consists of a problem of coupled damped and driven classical harmonic oscillators. 
Our goal is therefore to map Eq.\ref{eq:gen_q} to the linear charge density response of a quantum NP, so that its linear response polarization matches the macroscopic classical one. 
As shown previously in the simple DL dielectric function case\cite{fregoni2021}, such quantization approach corresponds to a macroscopic-QED 
quantization of the electromagnetic fields in the same dielectric environment\cite{fregoni2021,neuman2018,trugler2008}.

To this end it is convenient to recast Eq.\ref{eq:f_fit} as
\begin{eqnarray}\label{eq:f_fit2}
&f(\omega)&\approx\sum_p^N \frac{A_p}{\omega_p^2-\omega^2-i\gamma_p\omega}=\sum_p^N \frac{A_p}{2\overline{\omega}_p}\left(\frac{1}{\overline{\omega}_p-\omega-i\gamma_p/2}+\frac{1}{\overline{\omega}_p+\omega+i\gamma_p/2}\right)
\end{eqnarray}
where $\overline{\omega}_p=\sqrt{\omega_p^2-\gamma_p^2/4}$. 
Eq.\ref{eq:f_fit2} allows us to explicitly deal with the resonant and anti-resonant terms hidden in Eq.\ref{eq:f_fit}. 
This is a critical step for obtaining a set of coupled quantum oscillators representing the exact (classical) polarization response of the macroscopic system. 
Indeed, on the basis of Eq.\,\ref{eq:f_fit2}, the response charges can be decomposed as $\bm{q}(\omega)=\sum_p^N\bm{q}^R_p(\omega)+\bm{q}^A_p(\omega)$, where $\bm{q}_p^R(\omega),\bm{q}_p^A(\omega)$ respectively arise from the resonant and anti-resonant terms of Eq.\ref{eq:f_fit2}. 
Similarly to the DL case\cite{fregoni2021}, we now express the PCM-NP kernel in its diagonal form via the eigenmode expansion $(\bm{S}^{-1/2}\bm{DAS}^{1/2}=\bm{T}\bm{\lambda}\bm{T}^{\dagger}$, note that only the geometry of the NP, not the nature of the material is involved in this step). 
To clarify the meaning of such eigenmodes, it is useful to note that for a sphere they correspond to surface charge distributions with different multipolar characters (dipole, quadrupole, etc.). 
Upon using such eigenmode expansion and the decomposition of Eq.\ref{eq:f_fit2}, Eq.\ref{eq:gen_q} can be recast as (SI 1),
\begin{eqnarray}\label{eq:gen_q2}
&& \left(\bm{S}^{1/2}\bm{T}\bm{K}_{pp}^{R/A}(\omega)\bm{T}^{\dagger}\bm{S}^{1/2}\right)\bm{q}_p^{R/A}(\omega)+ \sum_{p' \neq p}^{N}\left(\bm{S}^{1/2}\bm{T}\bm{\tilde{\lambda}}_{pp'}\bm{T}^{\dagger}\bm{S}^{1/2}\right)\bm{q}_{p'}^{R/A}(\omega)+ \nonumber \\ 
&& \sum_{p'}^{N}\left(\bm{S}^{1/2}\bm{T}\bm{\tilde{\lambda}}_{pp'}\bm{T}^{\dagger}\bm{S}^{1/2}\right)\bm{q}_{p'}^{A/R}(\omega)= -\bm{V}(\omega)
\end{eqnarray}
where $\bm{K}_{pp}^R(\omega)$, $\bm{K}_{pp}^A(\omega)$ and $\bm{\tilde{\lambda}}_{pp'}$ are diagonal matrices on the surface element eigenmode index $\theta$: 
\begin{eqnarray}
&& \left(K_{pp}^R(\omega)\right)_{\theta \theta}=\frac{\frac{4\pi\overline{\omega}_p}{A_p}(\overline{\omega}_p-\omega-i\frac{\gamma_p}{2})+\lambda_\theta}{2\pi+\lambda_\theta} \\
&& \left(K_{pp}^A(\omega)\right)_{\theta \theta}=\frac{\frac{4\pi\overline{\omega}_p}{A_p}(\overline{\omega}_p+\omega+i\frac{\gamma_p}{2})+\lambda_\theta}{2\pi+\lambda_\theta} \\
&& \left(\tilde{\lambda}_{pp'}\right)_{\theta \theta}=\frac{\lambda_\theta}{2\pi+\lambda_\theta} .
\label{eq:gen_q_el}
\end{eqnarray}

Further manipulation of Eqs.\ref{eq:gen_q2}-\ref{eq:gen_q_el} (see SI 1), leads to independent PCM-NP response equations for each $\theta$th BEM eigenmode, reading
\begin{equation}\label{eq:mat_cl}
\left[
\begin{pmatrix}
\mathbb{A_\theta} & \mathbb{B_\theta} \\
\mathbb{B}_\theta^* & \mathbb{A}_\theta^*
\end{pmatrix}
-\omega
\begin{pmatrix}
\mathbb{I} & 0 \\
0 & \mathbb{-I}
\end{pmatrix}
\right]
\begin{pmatrix}
\mathbb{q}_\theta^R(\omega) \\
\mathbb{q}_\theta^A(-\omega)
\end{pmatrix}
=-
\begin{pmatrix}
\mathbb{V}_\theta (\omega)\\
\mathbb{V}_\theta^*(-\omega)
\end{pmatrix}
\end{equation}
where $\mathbb{A}_{\theta}\, ,\mathbb{B}_{\theta}$ are N$\times$N matrices, whereas  $\mathbb{q}_\theta^{R/A}\, , \mathbb{V}_\theta$ are N-dimensional vectors whose elements read
\begin{eqnarray}
&&(\mathbb{A}_{\theta})_{pp'}=\left(\overline{\omega}_p-i\frac{\gamma_p}{2}+\lambda_\theta\frac{A_p}{4\pi\overline{\omega}_p} \right)\delta_{pp'}+\left(1-\delta_{pp'}\right)\sqrt{\frac{A_p}{2\overline{\omega}_p}}\frac{\lambda_\theta}{2\pi}\sqrt{\frac{A_{p'}}{2\overline{\omega}_{p'}}} \\
&&(\mathbb{B}_{\theta})_{pp'}=\sqrt{\frac{A_p}{2\overline{\omega}_p}}\frac{\lambda_\theta}{2\pi}\left(\sqrt{\frac{A_{p'}}{2\overline{\omega}_{p'}}}\right)^* \\ 
&&\mathbb{q}_{\theta,p}^{R}(\omega)=\frac{1}{\sqrt{\frac{A_p}{2\overline{\omega}_p}\left(1+\frac{\lambda_\theta}{2\pi}\right)}}\sum_k\left(T^{\dagger}S^{1/2}\right)_{\theta k}q_{p,k}^{R}(\omega) \\\
&&\mathbb{q}_{\theta,p}^{A}(-\omega)=\frac{1}{\left(\sqrt{\frac{A_p}{2\overline{\omega}_p}\left(1+\frac{\lambda_\theta}{2\pi}\right)}\right)^*}\sum_k\left(T^{\dagger}S^{1/2}\right)_{\theta k}q_{p,k}^{A}(\omega) \\
&&\mathbb{V}_{\theta, p}(\omega)=\text{sgn}(A_p)\left(\sqrt{\frac{A_p}{2\overline{\omega}_p}\left(1+\frac{\lambda_\theta}{2\pi}\right)}\right)^*\sum_k\left(T^{\dagger}S^{-1/2}\right)_{\theta k}V_k(\omega) \\
&&\mathbb{V}_{\theta, p}^*(-\omega)=\text{sgn}(A_p)\sqrt{\frac{A_p}{2\overline{\omega}_p}\left(1+\frac{\lambda_\theta}{2\pi}\right)}\sum_k\left(T^{\dagger}S^{-1/2}\right)_{\theta k}V_k(\omega)\, .
\label{eq:mat_el}
\end{eqnarray}
Further details on the matrices $\mathbb{A}_{\theta}\, ,\mathbb{B}_{\theta}$ as well as a graphical representation of their structure is given in SI 1.

Notably, the shape of the classical PCM-NP response equations for each $\theta$th BEM eigenmode (Eq.\ref{eq:mat_cl}) strongly resembles the shape of the linear response equation of a quantum system (as in time-dependent density functional theory)\cite{mcweeny,corni2002,norman2018,ye2024},
\begin{equation}\label{eq:mat_q}
\left[
\begin{pmatrix}
\mathbb{A} & \mathbb{B} \\
\mathbb{B}^* & \mathbb{A}^*
\end{pmatrix}
-\omega
\begin{pmatrix}
\mathbb{I} & 0 \\
0 & \mathbb{-I}
\end{pmatrix}
\right]
\begin{pmatrix}
\mathbb{X}(\omega) \\
\mathbb{Y}(-\omega)
\end{pmatrix}
=-
\begin{pmatrix}
\mathbb{{V}}(\omega) \\
\mathbb{V}^*(-\omega)
\end{pmatrix}
\end{equation}
where the matrix $\mathbb{A}$ usually contains single particle excitation frequencies along the diagonal and couplings among them in the off-diagonal blocks, whereas $\mathbb{B}$ couples excitations and de-excitations. 
$\mathbb{X}(\omega),\mathbb{Y}(-\omega)$ respectively contain the Fourier transformed resonant and anti-resonant transition amplitudes describing the first-order change in the system density matrix upon perturbation, while $\mathbb{V}(\omega)$  and $\mathbb{V}^*(-\omega)$ store matrix elements of the perturbation. 

By comparing Eq.\ref{eq:mat_cl} with Eq.\ref{eq:mat_q} we can achieve quantization of the classical PCM-NP model with generic dielectric function. 
This is done by mapping the $\mathbb{A}_{\theta}$ and $\mathbb{B}_{\theta}$ matrices to the $\mathbb{A}$ and $\mathbb{B}$ blocks of the linear response matrix of a quantum system featuring single particle plasmonic excitation frequencies $\overline{\omega}_p+\lambda_\theta\frac{A_p}{4\pi\overline{\omega}_p}$ and damping rates $\gamma_p/2$. In turn, $\mathbb{V}_\theta(\omega)$ can be identified with the external perturbation and therefore $\mathbb{q}_\theta^R(\omega)$ with $\mathbb{X}(\omega)$ and $\mathbb{q}_\theta^A(-\omega)$ with $\mathbb{Y} (-\omega)$.  
In other words, the classical macroscopic polarization equation (Eq.\ref{eq:pcm_np}) of the NP can be exactly mapped to the linear response polarization of a quantum system composed of a set of coupled and damped quantum plasmon modes for each $\theta$th geometric BEM eigenmode of the NP, independently.

By following standard response theory\cite{mcweeny,norman2018} (SI 1 for details), for each $\theta$ we can solve the generalized eigenvalue problem associated with Eq.\ref{eq:mat_q}:
\begin{equation}\label{eq:mat_res_a}
\begin{pmatrix}
\mathbb{A_\theta} & \mathbb{B_\theta} \\
\mathbb{B}_\theta^* & \mathbb{A}_\theta^*
\end{pmatrix}
\bm{U}_{\theta}
=\begin{pmatrix}
\mathbb{I} & 0 \\
0 & \mathbb{-I}
\end{pmatrix}
\bm{U}_{\theta}\bm{d}_{\theta}
\end{equation}
where $\bm{d}_{\theta}$ is the diagonal eigenvalue matrix with elements $d_{\theta,n}=\omega_{\theta,n}-i\frac{\gamma_{\theta,n}}{2}$ and $d_{\theta,-n}=-\omega_{\theta,n}-i\frac{\gamma_{\theta,n}}{2}$ for resonant and anti-resonant transitions, respectively. 
$\bm{U}_{\theta}$ collects in its columns the corresponding generalized eigenvectors. 
Their properties are recalled in SI 1. 
From Eq.\ref{eq:mat_res_a} we can identify $\omega_{\theta, n}$ with the true excitation energy of the quantum NP plasmonic state $\ket{\theta, n}$ with decay rate $\frac{\gamma_{\theta,n}}{2}$ for each $\theta$ BEM eigenmode . 
This leads to the following NP plasmonic Hamiltonian,
\begin{equation}\label{eq:A_H}
    \hat{H}_{NP}=\sum_{\theta,n}\left(\omega_{\theta, n}-i\frac{\gamma_{\theta,n}}{2}\right)\hat{b}^{\dagger}_{\theta,n}\hat{b}_{\theta,n}
\end{equation}
which can be coupled to any quantum chemistry molecular description as discussed below.

Moreover, we can also identify within this picture transition elements of relevant quantum operators.
Focusing on surface charges, similarly to Ref.\,\citenum{fregoni2021} upon introducing the quantized surface charge operator for the k-th surface tessera $\hat{q}_k$, the quantum transition charges for each coupled $\ket{\theta,p}$ state can be identified:
\begin{equation}\label{eq:q_trans}
\bra{0}\hat{q}_k\ket{\theta,p}=\left(S^{-1/2}T\right)_{k \theta}\sqrt{\frac{A_p}{2\overline{\omega}_p}\left(1+\frac{\lambda_\theta}{2\pi}\right)}
\end{equation}
leading in turn to
\begin{equation}\label{eq:q_trans_n}
\bra{0}\hat{q}_k\ket{\theta,n}=\sum_{p}\left(S^{-1/2}T\right)_{k \theta}\left[\sqrt{\frac{A_p}{2\overline{\omega}_p}\left(1+\frac{\lambda_\theta}{2\pi}\right)}X_{\theta,pn} + \left(\sqrt{\frac{A_p}{2\overline{\omega}_p}\left(1+\frac{\lambda_\theta}{2\pi}\right)}\right)^*Y_{\theta,pn} \right] \, .
\end{equation}
The quantities defined in Eqs.\ref{eq:q_trans}-\ref{eq:q_trans_n} are transition matrix elements of the surface charge operator $\hat{q}$, which is a quantum operator in this formalism\cite{fregoni2020,guido2020}. 

Based on Eqs.\ref{eq:q_trans_n} and response theory\cite{mcweeny,norman2018} (SI 1 for details), the dipole-dipole polarizability tensor $\alpha_{ab}(\omega)$ of the quantum NP can be expressed in its spectral form as 
\begin{equation}\label{eq:pol}
 \alpha_{ab}(\omega)=\sum_{\theta,n} \frac{\bra{0}\hat{\mu}_a\ket{\theta,n}\bra{\theta,n}\hat{\mu}_b\ket{0}}{\omega_{\theta, n}-\omega-i\frac{\gamma_{\theta,n}}{2}}+ \\
 \frac{\bra{0}\hat{\mu}_a\ket{\theta,n}^*\bra{\theta,n}\hat{\mu}_b\ket{0}^*}{\omega_{\theta,n}+\omega+i\frac{\gamma_{\theta,n}}{2}}
\end{equation}
where 
 $\hat{\mu}_a$ is defined as $\hat{\mu}_a=\sum_k\hat{q}_k\vec{r}_{k,a}$ with $\vec{r}_{k,a}$ being the $a$th component of the position vector pointing to the $k$th NP tessera. 
 Eq.\ref{eq:pol} holds for all positive $A_p$. 
 A slightly more complex expression is obtained in the general case (see SI).

We stress that the plasmonic states so introduced should be regarded as effective states that describe the more complex electronic structure of the true metal NP, featuring quasi-continuum bands, and that they do not constitute exact eigenstates of any unperturbed Hamiltonian. 
If this were the case, the $\mathbb{B}_\theta$ block would be exactly zero according to response theory from exact states\cite{norman2018}. 
This is clearly not the case for a finite NP, where $\mathbb{B}_\theta$ is non-vanishing. 
Interestingly, for an infinitely planar plasmonic surface $\lambda_\theta$ would be zero\cite{corni2015}, making the coupling blocks vanishing.

\begin{figure}[tb]
\centering
\includegraphics{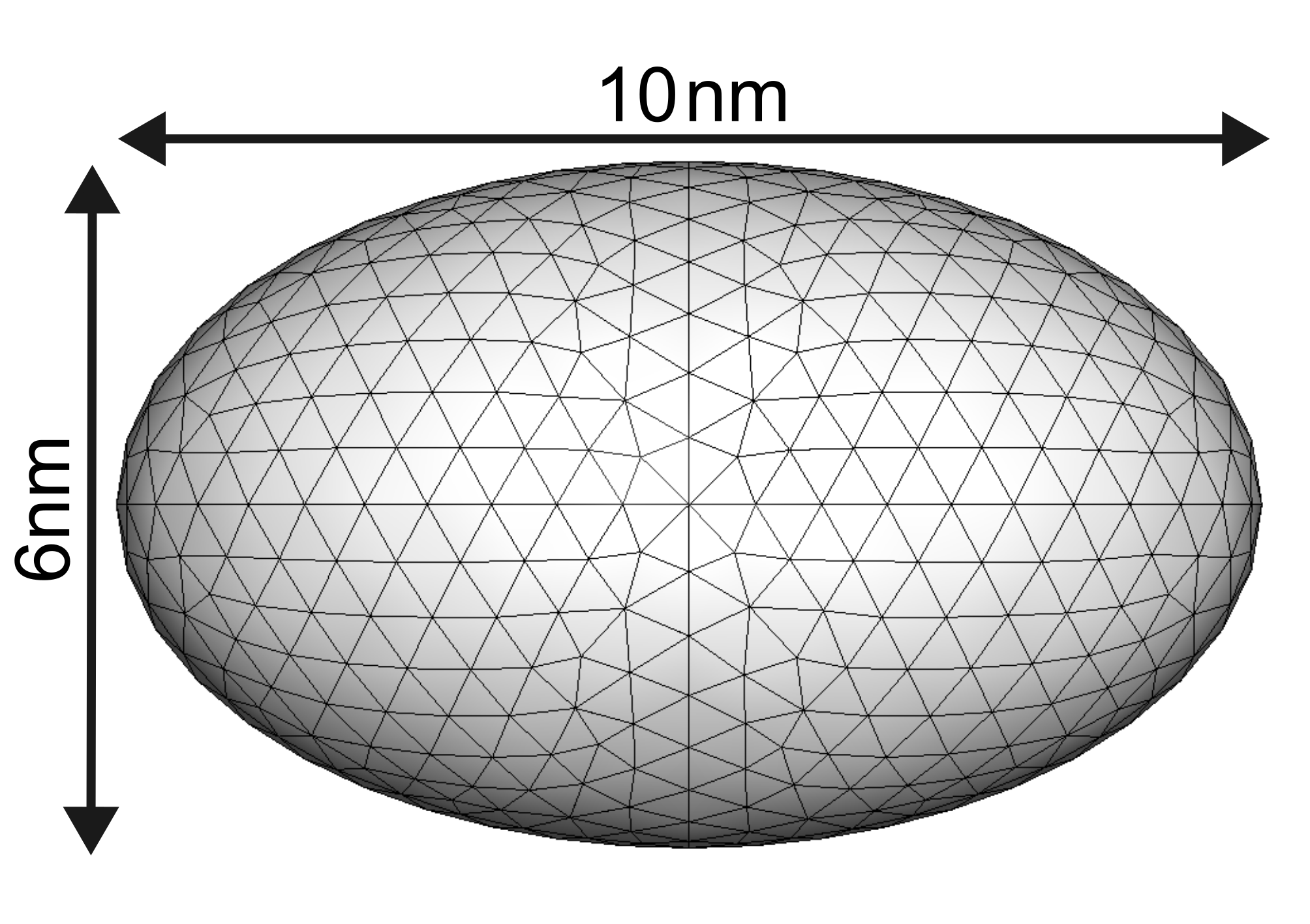}
\caption{ Ellipsoidal NP model used for simulations featuring 1371 surface tesserae.}
\label{fig:NP}
\end{figure}

\begin{figure*}[tb]
\centering
\includegraphics[width=\textwidth]{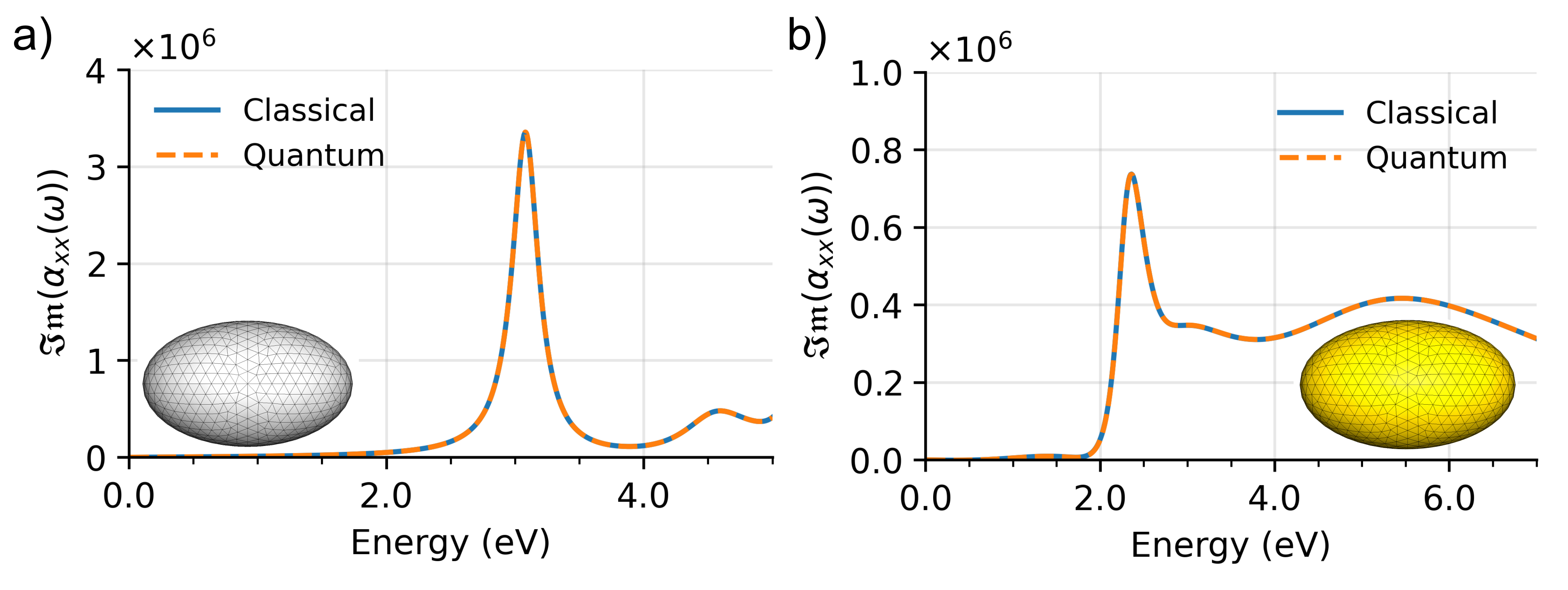}
\caption{Imaginary part of the xx component of the dipole-dipole polarizability tensor for the ellipsoidal NP (Fig.\,\ref{fig:NP}) when the $f(\omega)$ function is fitted to Ag Brendel-Bormann\cite{rakic1998} (a) or Au Etchegoin\cite{etchegoin2006} (b) reference data. The classical PCM-NP (solid blue) and quantum Q-PCM-NP (dashed orange) results are shown.}
\label{fig:AgAu_im}
\end{figure*}

The theory presented above has been numerically validated on a test case NP of ellipsoidal shape (Fig.\ref{fig:NP}) either made of silver or gold.
These two widely used plasmonic metals are well-known to exhibit multiple interband transitions that cannot be captured by a simple DL dielectric function model, thereby making a generic dielectric function approach desirable in these cases. 
The corresponding function $f(\omega)$ (Eq.\ref{eq:f_fit}) has been fitted to data from Ref.\citenum{rakic1998} for Ag and Ref.\citenum{etchegoin2006} for Au. 
The fitted pole parameters are reported in the computational details section (SI 3).
In Fig.\ref{fig:AgAu_im} the imaginary part of the NP polarizability, which is proportional to the NP absorption cross-section, is shown for both silver (Fig.\ref{fig:AgAu_im}a) and gold (Fig.\ref{fig:AgAu_im}b). 
The classical result (solid blue) is obtained from the induced dipole moment of the NP upon numerical solution of Eq.\ref{eq:pcm_np}, while the quantum result (dashed orange) is computed according to Eq.\ref{eq:pol}. 
Since in the quasi-static limit only the dipolar mode is contributing to the NP optical response only that mode is considered here (SI 3). 
The driving field is oriented along the main axis of the ellipsoidal NP (x axis) and so the $\alpha_{xx}(\omega)$ component of the polarizability tensor is analyzed. 
Fig.\ref{fig:AgAu_im} clearly shows that the here introduced quantum states of the NP correctly reproduce the classical macroscopic polarization of the NP in both cases, thus validating the quantization procedure detailed above. 
The real part of the polarizability is also correctly reproduced, as shown in Figs. S3. 
The implementation has also been tested on a gold spherical NP (Fig. S4), confirming the robustness of the method against different NP shapes. 

The quantum mode description introduced above can be straightforwardly coupled to a quantum-chemical molecular description by extending the procedure of Ref.\citenum{fregoni2020}, originally developed for the simpler DL case. 
Following Ref.\citenum{fregoni2020} and Eq. \ref{eq:A_H}, the full plasmon–molecule Hamiltonian describing the interaction between a molecule and the NP quantum modes derived from the general dielectric function can be written as (SI 2),
\begin{eqnarray}\label{eq:H_full}
\hat{H}_{\text{tot}}&=&\hat{H}_{\text{mol}}+\hat{H}_{\text{NP}}+\hat{H}_{\text{int}}  \\
&=& \hat{H}_{\text{mol}}+ \sum_{\theta,n}\left(\omega_{\theta, n}-i\frac{\gamma_{\theta,n}}{2}\right)\hat{b}^{\dagger}_{\theta,n}\hat{b}_{\theta,n}+
\sum_{\theta,n,k}\hat{V}_k\left( \bra{\theta,n}\hat{q}_k\ket{0}\hat{b}^{\dagger}_{\theta,n}+  \bra{0}\hat{q}_k\ket{\theta,n}\hat{b}\right) \nonumber
\end{eqnarray}
where $\hat{H}_{\text{mol}}$ is the standard molecular electronic Hamitonian and $\hat{V}_k$ is the molecular electrostatic potential operator evaluated at the centroid of the $k\text{-th}$ surface tessera of the NP.

The complete Hamiltonian in Eq.\ref{eq:H_full} can, in principle, be solved fully ab initio using QED–quantum chemistry methods to capture electron–plasmon correlation effects, as demonstrated in Ref. \citenum{fregoni2020}. 
However, this approach is often computationally demanding due to the additional plasmonic degrees of freedom beyond the already costly electronic structure treatment. 
A viable strategy to reduce this cost is to first calculate a given number of electronic states of the isolated molecule and subsequently couple them to the plasmonic modes, as in Ref.\citenum{romanelli2024}. 
This approximation is generally justified when the plasmon–molecule interaction is not particularly strong and so plasmon-molecule correlation effects remain limited.
In the present work we follow such approximation.
Therefore, starting from Eq.\ref{eq:H_full} and considering the molecular eigenstates, $\ket{g} \text{(for the ground state)}$ and $\ket{e} \text{(for a general excited state)}$,  $\hat{H}_{\text{tot}}$ in the rotating wave approximation can be expressed as (SI 2 for details)
\begin{eqnarray}\label{eq:H_full_matrix}
\hat{H}_{\text{tot}}=&&\omega_g\ket{g}\bra{g}+\sum_e\left(\omega_e-i\frac{\gamma_e}{2}\right)\ket{e}\bra{e}+\sum_{\theta,n}\left(\omega_{\theta, n}-i\frac{\gamma_{\theta,n}}{2}\right)\hat{b}^{\dagger}_{\theta,n}\hat{b}_{\theta,n}+\nonumber \\
&&+\sum_{\theta,n,k}\left(\hat{V}_k^{gg}\ket{g}\bra{g}+\sum_e\hat{V}_k^{ee}\ket{e}\bra{e}\right)\left( \bra{\theta,n}\hat{q}_k\ket{0}\hat{b}^{\dagger}_{\theta,n}+  \bra{0}\hat{q}_k\ket{\theta,n}\hat{b}_{\theta,n}\right) \nonumber \\
&& +\sum_{\theta,n,k,e}\hat{V}_{k}^{eg}\left( \bra{\theta,n}\hat{q}_k\ket{0}\hat{b}^{\dagger}_{\theta,n}\ket{g}\bra{e}+\bra{0}\hat{q}_k\ket{\theta,n}\hat{b}_{\theta,n}\ket{e}\bra{g}\right) \nonumber \\
&& +\sum_{\theta,n,k,e<e'}\hat{V}_{k}^{ee'}\left( \bra{\theta,n}\hat{q}_k\ket{0}\hat{b}^{\dagger}_{\theta,n}\ket{e}\bra{e'}+\bra{0}\hat{q}_k\ket{\theta,n}\hat{b}_{\theta,n}\ket{e'}\bra{e}\right) 
\, .
\end{eqnarray}

\begin{figure*}[tb]
\centering
\includegraphics[width=\textwidth]{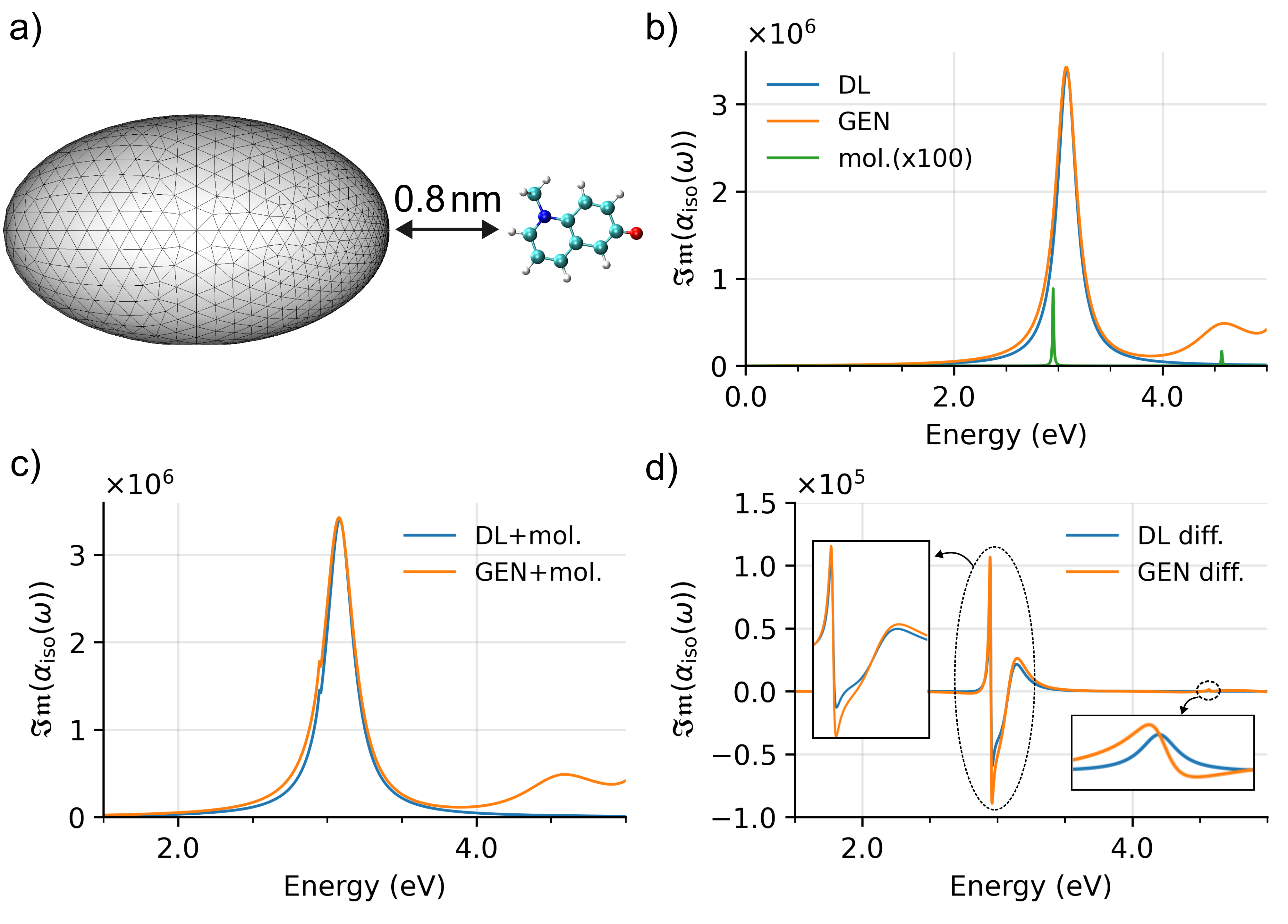}
\caption{a) Setup used for investigating the quantum coupling between an ellipsoidal silver
NP and one molecule (N-methyl-6-quinolone) described at ab initio level. 
The molecule-NP dimensions are not to scale. 
The NP dimensions are the same of Fig.\ref{fig:NP} but more surface tesserae are considered to refine the region closer to the molecule. Overall 2122 surface tesserae are used. b) Imaginary part of the isotropic dipole-dipole polarizability for the ellipsoidal NP of panel a) when the $f(\omega)$ function is fitted to Ag Brendel-Bormann\cite{rakic1998} reference data (orange) or when a simpler DL model is used (blue). The DL parameters are chosen to  reproduce the main plasmonic peak around 3.0 eV. The green line is the isotropic polarizability of the bare molecule scaled by a factor of 100. c) Corresponding polarizability plots of the coupled system (NP + molecule) when the generic dielectric function (orange) or DL quantum modes (blue) are used to investigate the molecule-NP coupling. d) Differences between the isotropic polarizability of the coupled system (panel c) and the corresponding bare plasmonic contribution shown in panel b for the two cases.}
\label{fig:coupled}
\end{figure*}

In Eq.\ref{eq:H_full_matrix} $\omega_g$ and $\omega_e$ are the molecular energies of ground and excited states, respectively, while $\gamma_e$ is a phenomenological damping rate to account for the finite lifetime of the molecular excited states.
Furthermore, $\hat{V}_k^{ee}$ and $\hat{V}_k^{eg}$ are shorthand notations for the matrix elements $\bra{g}\hat{V}_k\ket{g}$, $\bra{e}\hat{V}_k\ket{e}$ and $\bra{e}\hat{V}_k\ket{g}$. 
The first term accounts for NP-molecule ground-state polarization effects, the second for NP-molecule excited-state polarization effects whereas the third term $\hat{V}_k^{eg}$, which is the electrostatic potential on the NP surface due to the molecular transition $\ket{g}\rightarrow\ket{e}$, is responsible for Jaynes-Cummings-like coupling terms.  Additionally, the terms present in the last line of Eq.\ref{eq:H_full_matrix} promote a mixing of the bare molecular excited states due to molecule-NP polarization effects.

In the present work the Hamiltonian of Eq.\ref{eq:H_full_matrix} is diagonalized in the single-excitation subspace, $\ket{g}\otimes\ket{0_{\theta,n}},\ket{e}\otimes\ket{0_{\theta,n}},\ket{g}\otimes\ket{1_{\theta,n}}$ where the notation $\ket{g}\otimes\ket{1_{\theta,n}}$ means that the molecule is in its ground-state but the plasmon mode $\ket{\theta,n}$ is excited.

Diagonalization of the resulting non-Hermitian Hamiltonian leads to the dipole–dipole polarizability of the coupled molecule–NP system, analogously to Eq.\ref{eq:pol}.
The real and imaginary parts of the eigenvalues of $\hat{H}_{\text{tot}}$ correspond to the absorption energies and damping rates of the plexcitonic states, respectively. 
The associated transition dipoles can be constructed from the left and right eigenvectors of $\hat{H}_{\text{tot}}$, using the molecular transition dipoles and those of the Q-PCM-NP modes (SI 1–2).

Such methodology has been tested to study the plasmon-molecule coupling effects between a silver ellipsoidal NP and a quinolone molecule described at CIS/6-31g* level of theory, using a setup similar to that of Ref.\citenum{romanelli2024}. 
The molecule is placed 0.8 nm far from the NP surface, as shown in Fig.\ref{fig:coupled}a. 
The ellipsoidal NP has the same size and Ag dielectric function as the model in Figs.\ref{fig:NP}–\ref{fig:AgAu_im}a), but employs a refined surface tessellation in the region closer to the molecule to improve the numerical evaluation of plasmon–molecule coupling terms. 
The imaginary part of the isotropic polarizability $\alpha_{\text{iso}}=\frac{1}{3}(\alpha_{xx}+\alpha_{yy}+\alpha_{zz})$ of the coupled system is investigated (Fig.\ref{fig:coupled}b-d) using either the general dielectric function (orange) or a single DL oscillator (blue), the latter relying on the simpler Q-PCM-NP scheme detailed in Ref.\citenum{fregoni2021}. The DL parameters (SI 3) are tuned to reproduce the main plasmonic peak at $\approx 3.0\,\text{eV}$.
As shown by the isolated components (Fig.\ref{fig:coupled}a), the first two excited states of the quinolone molecule absorb at $\approx 2.95\,\text{eV}$ and $\approx 4.57\,\text{eV}$ and they are almost resonating with the two plasmonic transitions predicted by the generic dielectric function (Fig.\ref{fig:coupled}b). 
Furthermore, the molecular transitions are considerably weaker than the plasmonic ones, approximately $40\,\degree$ tilted with respect to the NP main axis and the plasmon-molecule coupling between the brighest plasmon and the lowest excited-state of quinolone is $\approx\,8\text{meV}$, thereby not particularly strong.

In the spectral region around 3.0 eV, where the two dielectric models show minimal differences, only minor variations are observed in the full-system spectrum (Fig.\ref{fig:coupled}c). 
These differences become more evident upon subtracting the corresponding bare plasmonic contribution from the polarizability of the coupled (molecule + NP) system, yielding the differential polarizability shown in Fig.\ref{fig:coupled}d, where the molecular contribution is evidenced.
This representation reveals that near the plasmonic peak at 3.0 eV, both the DL and generic dielectric function models exhibit a similar sigmoidal-like behavior, consistent with plasmon–molecule coupling and so mixing. 
In contrast, significant discrepancies emerge around the second plasmonic peak at $4.5\,\text{eV}$. 
Here, the DL model displays a simple Lorentzian peak (blue line, see close-up), indicative of negligible mixing, whereas the generic dielectric function (orange line) retains a weaker yet distinct sigmoidal feature, pointing to significant coupling and mixing with the second excited state of quinolone.
This behavior is consistent with the plasmonic spectra of the two models (Fig.\ref{fig:coupled}b) and highlights a key limitation of the DL approach: its inherent restriction to describing a single plasmonic resonance at a time. 
As a result, it becomes inadequate when multiple plasmonic modes contribute to the overall optical response of the coupled system.

Additionally, it is worth emphasizing that the parameter tuning required to reproduce a given plasmonic peak (Fig.\ref{fig:coupled}a) with DL dielectric function is system-specific and can be cumbersome, limiting the transferability of the approach. 
In contrast, the Q-PCM-NP framework based on a generic dielectric function provides a more robust and broadly applicable description of the plasmonic response, enabling a consistent treatment across different systems and spectral regions with no need to system-specific tuning procedures.

The Q-PCM-NP scheme with generic dielectric function introduced in this work can be readily applied to more complex plasmonic structures, such as scanning tunneling microscope tips and NP aggregates, previously treated within the simpler DL framework.\cite{romanelli2023,fregoni2021} 
The underlying theoretical framework remains unchanged.

In this work, we have laid down an effective quantum modes description to model the optical response of arbitrarily-shaped NPs endowed with empirical dielectric functions. 
The linear response polarization of the quantum NP correctly recovers the classical result obtained by solving the macroscopic Maxwell equations in the quasistatic limit via BEM. 
It is worth remarking that although the NP linear response is usually well-described by classical electrodynamics, the quantization of the NP response is of utmost importance when the coupling with an external quantum emitter (i.e. molecule) is investigated. 
Indeed, the classical model is known to provide faulty behaviors both in the strong-coupling and strong-driving regimes\cite{romanelli2024,waks2010}, making it inadequate to realistically model plexcitonic systems. 
Since BEM has extensively proven to be a convenient framework to model ab initio molecules in complex plasmonic environments\cite{neuman2018,mennucci2019,romanelli2023}, the development here presented paves the way for a fully quantum description of plexcitonic systems, composed of NPs of arbitrary shape and made of realistic metals.

\begin{acknowledgement}
The authors acknowledge Giulia Dall'Osto for insightful preliminary discussions on the generic dielectric function fitting procedure. The authors also acknowledge the C3P HPC facility of the Department of Chemical Sciences of the University of Padua for
providing the computational resources supporting this work.
\end{acknowledgement}

\begin{suppinfo}
Detailed derivation of the Q-PCM-NP quantum modes with a generic dielectric function;  Coupling to a quantum chemistry molecular description; Computational details; Real part of polarizability for the ellipsoidal NP of Fig.1; Results for a gold spherical NP.  

\end{suppinfo}

\bibliography{bib.bib}

\end{document}


\tableofcontents
\newpage
\section{Q-PCM-NP with generic dielectric function}\label{sec:SI_theory}

The goal of this section is to manipulate the classical PCM-NP equation so to arrive to a format that is identical to the response equation of a quantum system.
\newline Within the generic dielectric function approach\cite{dall2020real}, the classical PCM-NP equations read 
\begin{equation}\label{eq:SI_gen}
    \bm{q}(\omega)=\frac{1}{2\pi}f(\omega)\bm{F}(\omega)
\end{equation}
with
\begin{equation}\label{eq:SI_gen1}
    \bm{F}(\omega)=-\left[\bm{A}\bm{D}^{\dagger}\bm{q}(\omega)+\bm{S}^{-1}\left(2\pi\mathbb{I}+\bm{DA}\right)\bm{V}(\omega)\right]
\end{equation}
and
\begin{equation}\label{eq:SI_f_fit}
f(\omega)=\frac{\epsilon(\omega)-1}{\epsilon(\omega)+1}\approx\sum_p^N\frac{A_p}{\omega_p^2-\omega^2-i\gamma_p\omega}\,.
\end{equation}
As discussed in Ref.\cite{dall2020real}, using Eq.\ref{eq:SI_f_fit} the classical response charges of Eq.\ref{eq:SI_gen} can be decomposed as pole-dependent charges as $\bm{q}(\omega)=\sum_p^N\bm{q}_p(\omega)$, turning Eq.\ref{eq:SI_gen} into
\begin{equation}\label{eq:SI_gen2}
\frac{2\pi}{A_p}(\omega_p^2-\omega^2-i\gamma_p\omega)\bm{q}_p(\omega)=-\bigr[ \bm{A}\bm{D}^{\dagger}\bm{q}(\omega)+\bm{S}^{-1}\left(2\pi\mathbb{I}+\bm{DA}\right)\bm{V}(\omega) \bigr] .
\end{equation}
To achieve quantization (Q-PCM-NP) it is more convenient to explicitly express the resonant and anti-resonant contributions hidden in Eq.\ref{eq:SI_f_fit} as
\begin{equation}\label{eq:SI_f_fit2}
f(\omega)\approx\sum_p^N\frac{A_p}{\omega_p^2-\omega^2-i\gamma_p\omega}=\sum_p^N \frac{A_p}{2\overline{\omega}_p}\left(\frac{1}{\overline{\omega}_p-\omega-i\gamma_p/2}+\frac{1}{\overline{\omega}_p+\omega+i\gamma_p/2}\right)
\end{equation}
with 
\begin{equation}\label{eq:SI_f_fit3}
\overline{\omega}_p=\sqrt{\omega_p^2-\gamma_p^2/4}.
\end{equation}
The first term in round brackets of Eq.\ref{eq:SI_f_fit2} can be defined as the resonant term (R, $\propto (\overline{\omega}_p-\omega)^{-1}$) while the other one is the anti-resonant (A, $\propto (\overline{\omega}_p+\omega)^{-1}$). Consequently, the classical response charges can be further decomposed as $\bm{q}(\omega)=\sum_p^N\bm{q}_p(\omega)=\sum_p^N\bm{q}^R_p(\omega)+\bm{q}^A_p(\omega)$, which turns Eq.\ref{eq:SI_gen2} into two separate coupled equations for each pth pole,
\begin{equation}
\begin{split} \label{eq:SI_gen2_s}   
& \frac{4\pi\overline{\omega}_p}{A_p}(\overline{\omega}_p-\omega-i\frac{\gamma_p}{2})\bm{q}_p^R(\omega)=-\bigr[ \bm{A}\bm{D}^{\dagger}\sum_{p'}^N\left(\bm{q}_{p'}^R(\omega)+\bm{q}_{p'}^A(\omega)\right)+\bm{S}^{-1}\left(2\pi\mathbb{I}+\bm{DA}\right)\bm{V}(\omega) \bigr]\, , \\
& \frac{4\pi\overline{\omega}_p}{A_p}(\overline{\omega}_p+\omega+i\frac{\gamma_p}{2})\bm{q}_p^A(\omega)=-\bigr[ \bm{A}\bm{D}^{\dagger}\sum_{p'}^N\left(\bm{q}_{p'}^R(\omega)+\bm{q}_{p'}^A(\omega)\right)+\bm{S}^{-1}\left(2\pi\mathbb{I}+\bm{DA}\right)\bm{V}(\omega) \bigr] \, .
\end{split}
\end{equation}
Furthermore, by turning the BEM kernel in diagonal form as shown in Refs.\cite{fregoni2020,corni2015} via the eigenmode decomposition $(\bm{S}^{-1/2}\bm{DAS}^{1/2}=\bm{T}\bm{\lambda}\bm{T}^{\dagger})$ and by left-multiplying each Eq.\ref{eq:SI_gen2_s} by $\bm{S}$ (while considering the following relations\cite{corni2015} $\bm{S}=\bm{S}^{\frac{1}{2}}\bm{S}^{\frac{1}{2}}$, $\bm{SAD}^{\dagger}=\bm{DAS}$), straightforward matrix algebra leads to
\begin{equation}\label{eq:SI_gen3_s}
\begin{split}
\left(\bm{S}^{1/2}\bm{T}\bm{K}_{pp}^R(\omega)\bm{T}^{\dagger}\bm{S}^{1/2}\right)\bm{q}_p^R(\omega)+&
\sum_{p' \neq p}^{N}\left(\bm{S}^{1/2}\bm{T}\bm{\tilde{\lambda}}_{pp'}\bm{T}^{\dagger}\bm{S}^{1/2}\right)\bm{q}_{p'}^R(\omega)+ \\ 
& \sum_{p'}^{N}\left(\bm{S}^{1/2}\bm{T}\bm{\tilde{\lambda}}_{pp'}\bm{T}^{\dagger}\bm{S}^{1/2}\right)\bm{q}_{p'}^A(\omega)= -\bm{V}(\omega)\, , \\
\left(\bm{S}^{1/2}\bm{T}\bm{K}_{pp}^A(\omega)\bm{T}^{\dagger}\bm{S}^{1/2}\right)\bm{q}_p^A(\omega)+& \sum_{p' \neq p}^{N}\left(\bm{S}^{1/2}\bm{T}\bm{\tilde{\lambda}}_{pp'}\bm{T}^{\dagger}\bm{S}^{1/2}\right)\bm{q}_{p'}^A(\omega) + \\ 
& \sum_{p'}^{N} \left(\bm{S}^{1/2}\bm{T}\bm{\tilde{\lambda}}_{pp'}\bm{T}^{\dagger}\bm{S}^{1/2}\right)\bm{q}_{p'}^R(\omega)= -\bm{V}(\omega)
\end{split}
\end{equation}
with 
\begin{equation}\label{eq_SI_gen_el_s}
\begin{split}
& K_{pp,\theta\theta}^R(\omega)=\frac{\frac{4\pi\overline{\omega}_p}{A_p}(\overline{\omega}_p-\omega-i\frac{\gamma_p}{2})+\lambda_\theta}{2\pi+\lambda_\theta}=\frac{\overline{\omega}_p-\omega-i\frac{\gamma_p}{2}+\lambda_\theta\frac{A_p}{4\pi\overline{\omega}_p}}{\frac{A_p}{2\overline{\omega}_p}\left(1+\frac{\lambda_\theta}{2\pi}\right)}\, , \\
& K_{pp,\theta\theta}^A(\omega)=\frac{\frac{4\pi\overline{\omega}_p}{A_p}(\overline{\omega}_p+\omega+i\frac{\gamma_p}{2})+\lambda_\theta}{2\pi+\lambda_\theta}=\frac{\overline{\omega}_p+\omega+i\frac{\gamma_p}{2}+\lambda_\theta\frac{A_p}{4\pi\overline{\omega}_p}}{\frac{A_p}{2\overline{\omega}_p}\left(1+\frac{\lambda_\theta}{2\pi}\right)}\, , \\
& \tilde{\lambda}_{pp',\theta\theta}=\frac{\lambda_\theta}{2\pi+\lambda_\theta} \,.
\end{split}
\end{equation}
Note that the matrices of Eq.\ref{eq_SI_gen_el_s} are diagonal in $\theta\theta'$ and the absence of $pp'$ indexes on the right-hand side of $\tilde{\lambda}_{pp',\theta\theta}=\frac{\lambda_\theta}{2\pi+\lambda_\theta}$ means that these coupling matrix elements do not depend on the pole indexes, but only on the BEM eigenmode one ($\theta$). Note that since the matrix $\bm{S}^{-1/2}\bm{DAS}^{1/2}$ is real and symmetric, the $\lambda_\theta$ eigenvalues are also all real. 
\newline
Making explicit matrix multiplication, Eq.\ref{eq:SI_gen3_s} reads
\begin{equation}\label{eq:SI_gen4_s}
\begin{split}
&\sum_{\theta k}\left(S^{1/2}T\right)_{j \theta}K_{pp,\theta\theta}^R\left(T^{\dagger}S^{1/2}\right)_{\theta k}q_{pk}^R+
\sum_{\theta k}\sum_{p' \neq p}^{N}\left(S^{1/2}T\right)_{j \theta}\tilde{\lambda}_{pp',\theta\theta}\left(T^{\dagger}S^{1/2}\right)_{\theta k}q_{p'k}^R+ \\
& \sum_{\theta k}\sum_{p'}^{N}\left(S^{1/2}T\right)_{j \theta}\tilde{\lambda}_{pp',\theta\theta}\left(T^{\dagger}S^{1/2}\right)_{\theta k}q_{p'k}^A= -V_j\, , \\
&\sum_{\theta k}\left(S^{1/2}T\right)_{j \theta}K_{pp,\theta\theta}^A\left(T^{\dagger}S^{1/2}\right)_{\theta k}q_{pk}^A+
\sum_{\theta k}\sum_{p' \neq p}^{N}\left(S^{1/2}T\right)_{j \theta}\tilde{\lambda}_{pp',\theta\theta}\left(T^{\dagger}S^{1/2}\right)_{\theta k}q_{p'k} ^A+ \\ 
&\sum_{\theta k}\sum_{p'}^{N}\left(S^{1/2}T\right)_{j \theta}\tilde{\lambda}_{pp',\theta\theta}\left(T^{\dagger}S^{1/2}\right)_{\theta k}q_{p'k}^R= -V_j
\end{split}
\end{equation}
where the $(\omega)$ have been dropped to ease notation and $q_{pk}^{R/A}=q_{pk}^{R/A}(\omega)$ represents the $k$th surface classical resonant/anti-resonant response charge due to the $p$th pole. $V_j=V_j(\omega)$ is instead the external potential acting on the jth surface tessera. 

\noindent Upon defining $\tilde{q}_{\theta p}^R=\sum_k\left(T^{\dagger}S^{1/2}\right)_{\theta k}q_{pk}^{R}\ ,\ \tilde{q}_{\theta p}^A = \sum_k\left(T^{\dagger}S^{1/2}\right)_{\theta k}q_{pk}^{A}$ and $\tilde{V}_{\theta p}=\sum_j\left(T^{\dagger}S^{-1/2}\right)_{\theta j}V_j$ Eq.\ref{eq:SI_gen4_s} can be recast as
\begin{equation}\label{eq:SI_gen5_s}
\begin{split}
&K_{pp,\theta\theta}^R\tilde{q}_{\theta p}^R+
\sum_{p' \neq p}^{N}\tilde{\lambda}_{pp',\theta\theta}\tilde{q}_{ \theta p'}^R+ \sum_{p'}^{N}\tilde{\lambda}_{pp',\theta\theta}\tilde{q}_{\theta p'}^A= -\tilde{V}_{\theta p}\, , \\
&K_{pp,\theta\theta}^A\tilde{q}_{\theta p}^A+
\sum_{p' \neq p}^{N}\tilde{\lambda}_{pp',\theta\theta}\tilde{q}_{ \theta p'}^A+ \sum_{p'}^{N}\tilde{\lambda}_{pp',\theta\theta}\tilde{q}_{\theta p'}^R= -\tilde{V}_{\theta p}\, .
\end{split}
\end{equation}
Note that we have added a $p$ subscript to the potential $\tilde{V}$ to make the terms of the  equations uniform, but that $\tilde{V}_{\theta p}$ is actually independent from $p$. \newline Further manipulation of Eq.\ref{eq:SI_gen5_s} finally leads to 
\begin{equation}\label{eq:SI_gen6_s}
\begin{split}
&\left( \overline{\omega}_p-i\frac{\gamma_p}{2}+\lambda_\theta\frac{A_p}{4\pi\overline{\omega}_p}\right)\mathbb{q}_{\theta p}^R-\omega \mathbb{q}_{\theta p}^R + \sum_{p' \neq p}^{N}\sqrt{\frac{A_p}{2\overline{\omega}_p}\left(1+\frac{\lambda_\theta}{2\pi}\right)}\tilde{\lambda}_{pp',\theta\theta}\sqrt{\frac{A_{p'}}{2\overline{\omega}_{p'}}\left(1+\frac{\lambda_\theta}{2\pi}\right)}\mathbb{q}_{\theta p'}^R + \\
& \sum_{p'}^{N}\sqrt{\frac{A_p}{2\overline{\omega}_p}\left(1+\frac{\lambda_\theta}{2\pi}\right)}\tilde{\lambda}_{pp',\theta\theta}\left(\sqrt{\frac{A_{p'}}{2\overline{\omega}_{p'}}\left(1+\frac{\lambda_\theta}{2\pi}\right)}\right)^*\mathbb{q}_{\theta p'}^A= -\mathbb{V}_{\theta p}\, , \\
&\left( \overline{\omega}_p+i\frac{\gamma_p}{2}+\lambda_\theta\frac{A_p}{4\pi\overline{\omega}_p}\right)\mathbb{q}_{\theta p}^A+\omega \mathbb{q}_{\theta p}^A +  
\sum_{p' \neq p}^{N}\left(\sqrt{\frac{A_p}{2\overline{\omega}_p}\left(1+\frac{\lambda_\theta}{2\pi}\right)}\right)^*\tilde{\lambda}_{pp',\theta\theta}\left(\sqrt{\frac{A_{p'}}{2\overline{\omega}_{p'}}\left(1+\frac{\lambda_\theta}{2\pi}\right)}\right)^* \\ 
& \mathbb{q}_{\theta p'}^A + 
\sum_{p'}^{N}\left(\sqrt{\frac{A_p}{2\overline{\omega}_p}\left(1+\frac{\lambda_\theta}{2\pi}\right)}\right)^*\tilde{\lambda}_{pp',\theta\theta}\left(\sqrt{\frac{A_{p'}}{2\overline{\omega}_{p'}}\left(1+\frac{\lambda_\theta}{2\pi}\right)}\right)\mathbb{q}_{\theta p'}^R= -\mathbb{V}_{\theta p}^*\, , \\
\end{split}
\end{equation}
where 
\begin{equation}\label{eq:SI_gen7_s}
\begin{split}
& \mathbb{q}_{\theta p}^R= \frac{1}{\sqrt{\frac{A_p}{2\overline{\omega}_p}\left(1+\frac{\lambda_\theta}{2\pi}\right)}}\tilde{q}_{\theta p}^R \, \\
& \mathbb{q}_{\theta p}^A= \frac{1}{\left(\sqrt{\frac{A_p}{2\overline{\omega}_p}\left(1+\frac{\lambda_\theta}{2\pi}\right)}\right)^*}\tilde{q}_{\theta p}^A \, \\
&\mathbb{V}_{\theta p}=\text{sgn}(A_p)\left(\sqrt{\frac{A_p}{2\overline{\omega}_p}\left(1+\frac{\lambda_\theta}{2\pi}\right)}\right)^*\tilde{V}_{\theta p}\, \\
& \mathbb{V}_{\theta p}^*=\text{sgn}(A_p)\sqrt{\frac{A_p}{2\overline{\omega}_p}\left(1+\frac{\lambda_\theta}{2\pi}\right)}\tilde{V}_{\theta p} \, .
\end{split}
\end{equation}
The $\text{sgn}(A_p)$ in the last two equations is needed to take into account that some $A_p$ may be negative, as described and explained in ref\cite{dall2020real}. In practice, we are translating this into a modified perturbation whose matrix elements have a change of sign for those $\ket{\theta,p}$ that corresponds to negative $A_p$. 

Recalling the form of $\tilde{\lambda}_{pp',\theta\theta}$ (Eq.\ref{eq_SI_gen_el_s}) Eq.\ref{eq:SI_gen6_s} can be further simplified to
\begin{equation}\label{eq:SI_gen8_s}
\begin{split}
&\left( \overline{\omega}_p-i\frac{\gamma_p}{2}+\lambda_\theta\frac{A_p}{4\pi\overline{\omega}_p}\right)\mathbb{q}_{\theta p}^R-\omega \mathbb{q}_{\theta p}^R + \sum_{p' \neq p}^{N}\sqrt{\frac{A_p}{2\overline{\omega}_p}}\frac{\lambda_\theta}{2\pi}\sqrt{\frac{A_{p'}}{2\overline{\omega}_{p'}}}\mathbb{q}_{\theta p'}^R + \\
&\sum_{p'}^{N}\sqrt{\frac{A_p}{2\overline{\omega}_p}}\frac{\lambda_\theta}{2\pi}\left(\sqrt{\frac{A_{p'}}{2\overline{\omega}_{p'}}}\right)^*\mathbb{q}_{\theta p'}^A= -\mathbb{V}_{\theta p}\, , \\
&\left( \overline{\omega}_p+i\frac{\gamma_p}{2}+\lambda_\theta\frac{A_p}{4\pi\overline{\omega}_p}\right)\mathbb{q}_{\theta p}^A+\omega \mathbb{q}_{\theta p}^A +  
\sum_{p' \neq p}^{N}\left(\sqrt{\frac{A_p}{2\overline{\omega}_p}}\right)^*\frac{\lambda_\theta}{2\pi}\left(\sqrt{\frac{A_{p'}}{2\overline{\omega}_{p'}}}\right)^*\mathbb{q}_{\theta p'}^A + \\ 
& \sum_{p'}^{N}\left(\sqrt{\frac{A_p}{2\overline{\omega}_p}}\right)^*\frac{\lambda_\theta}{2\pi}\left(\sqrt{\frac{A_{p'}}{2\overline{\omega}_{p'}}}\right)\mathbb{q}_{\theta p'}^R= -\mathbb{V}_{\theta p}^*\, .
\end{split}
\end{equation}
Basically, for each $\theta$th BEM eigenmode Eq.\ref{eq:SI_gen8_s} corresponds to an independent matrix equation of the following form
\begin{equation}\label{eq:SI_mat_cl}
\left[
\begin{pmatrix}
\mathbb{A_\theta} & \mathbb{B_\theta} \\
\mathbb{B}_\theta^* & \mathbb{A}_\theta^*
\end{pmatrix}
-\omega
\begin{pmatrix}
\mathbb{I} & 0 \\
0 & \mathbb{-I}
\end{pmatrix}
\right]
\begin{pmatrix}
\mathbb{q}_\theta^R \\
\mathbb{q}_\theta^A
\end{pmatrix}
=-
\begin{pmatrix}
\mathbb{V}_\theta \\
\mathbb{V}_\theta^*
\end{pmatrix}
\end{equation}
with
\begin{equation}\label{eq:SI_mat_cl_el}
\begin{split}
& (\mathbb{A}_{\theta})_{pp'}=\left(\overline{\omega}_p-i\frac{\gamma_p}{2}+\lambda_\theta\frac{A_p}{4\pi\overline{\omega}_p} \right)\delta_{pp'}+\left(1-\delta_{pp'}\right)\sqrt{\frac{A_p}{2\overline{\omega}_p}}\frac{\lambda_\theta}{2\pi}\sqrt{\frac{A_{p'}}{2\overline{\omega}_{p'}}} \\
& (\mathbb{B}_{\theta})_{pp'}=\sqrt{\frac{A_p}{2\overline{\omega}_p}}\frac{\lambda_\theta}{2\pi}\left(\sqrt{\frac{A_{p'}}{2\overline{\omega}_{p'}}}\right)^*
\end{split}
\end{equation}
and $\mathbb{q}_\theta^{R/A}$,$\mathbb{V}_\theta$ are vectors of dimension N (n.\,of poles in Eq.\ref{eq:SI_f_fit}) which store the corresponding $p$th elements $\mathbb{q}_{\theta,p}^{R/A}$, $\mathbb{V}_{\theta,p}$. 

\noindent To give a graphical illustration of the matrices $\mathbb{A}_{\theta},\,\mathbb{B}_{\theta}$, in the hypothetical case of 3 poles (N=3), the matrices would look

\begin{equation}\label{eq_SI_mat_cl_exa}
\begin{split}
  & \mathbb{A}_{\theta}=
  \begin{pmatrix}   
 \overline{\overline{K}}_{11,\theta \theta} & \overline{\overline{\lambda}}_{12,\theta\theta}   & \overline{\overline{\lambda}}_{13,\theta\theta} \\
\overline{\overline{\lambda}}_{12,\theta\theta}   & 
\overline{\overline{K}}_{22,\theta \theta} & \overline{\overline{\lambda}}_{23,\theta\theta} \\
\overline{\overline{\lambda}}_{13,\theta\theta}   & 
\overline{\overline{\lambda}}_{23,\theta\theta}   & 
\overline{\overline{K}}_{33,\theta \theta}
\end{pmatrix} \\
  & \mathbb{B}_{\theta}=
  \begin{pmatrix}   
 \overline{\overline{\eta}}_{11,\theta\theta} & \overline{\overline{\eta}}_{12,\theta\theta}   & \overline{\overline{\eta}}_{13,\theta\theta} \\
 \overline{\overline{\eta}}_{12,\theta\theta}^* & \overline{\overline{\eta}}_{22,\theta\theta}   & \overline{\overline{\eta}}_{23,\theta\theta} \\
 \overline{\overline{\eta}}_{13,\theta\theta}^* & \overline{\overline{\eta}}_{23,\theta\theta}^*   & \overline{\overline{\eta}}_{33,\theta\theta} \\
\end{pmatrix}
\end{split}
\end{equation}

\noindent with $\overline{\overline{K}}_{pp,\theta \theta}=\left(\overline{\omega}_p-i\frac{\gamma_p}{2}+\lambda_\theta\frac{A_p}{4\pi\overline{\omega}_p} \right)$, $\overline{\overline{\lambda}}_{pp', \theta \theta}=\sqrt{\frac{A_p}{2\overline{\omega}_p}}\frac{\lambda_\theta}{2\pi}\sqrt{\frac{A_{p'}}{2\overline{\omega}_{p'}}}$ and $\overline{\overline{\eta}}_{pp', \theta \theta}=\sqrt{\frac{A_p}{2\overline{\omega}_p}}\frac{\lambda_\theta}{2\pi}\left(\sqrt{\frac{A_{p'}}{2\overline{\omega}_{p'}}}\right)^*$ . Note that $\mathbb{A_{\theta}}$ is not hermitian because of the imaginary damping rates, whereas $\mathbb{B_{\theta}}$ is.

Notably, the form of Eq.\ref{eq:SI_mat_cl} bears a strong similarity with the matrix structure of the linear response equation of a quantum system. Indeed, standard linear response theory\cite{norman2018,ye2024,mcweeny} leads to the following general matrix equation
\begin{equation}\label{eq:SI_mat_q}
\left[
\begin{pmatrix}
\mathbb{A} & \mathbb{B} \\
\mathbb{B}^* & \mathbb{A}^*
\end{pmatrix}
-\omega
\begin{pmatrix}
\mathbb{I} & 0 \\
0 & \mathbb{-I}
\end{pmatrix}
\right]
\begin{pmatrix}
\mathbb{X} \\
\mathbb{Y}
\end{pmatrix}
=-
\begin{pmatrix}
\mathbb{{V}} \\
\mathbb{V}^*
\end{pmatrix}
\end{equation}
where the matrices $\mathbb{A},\mathbb{B}$ store single particle excitations of the system and couplings among them, whereas $\mathbb{X},\mathbb{Y}$ respectively contains the Fourier transformed resonant and anti-resonant transition amplitudes describing the first-order change in the system density matrix upon perturbation. $\mathbb{V}$  and $\mathbb{V}^*$ respectively store matrix elements of the perturbation. 

By comparing Eq.\ref{eq:SI_mat_cl} and Eq.\ref{eq:SI_mat_q}, it can be observed that the classical PCM-NP equations in the generic dielectric function approach can be exactly mapped to the linear response equations of a quantum system for each $\theta$th BEM eigenmode independently, where
\begin{equation}\label{eq:SI_map}
\begin{split}
& \mathbb{A}=\mathbb{A}_{\theta} \\
&\mathbb{B}=\mathbb{B}_{\theta} \\
& \mathbb{X}=\mathbb{q}_\theta^{R}\\
& \mathbb{Y}=\mathbb{q}_\theta^{A} \\
& \mathbb{V}=\mathbb{{V}_\theta} \, .\\
\end{split}
\end{equation}
Following this mapping, we can identify the diagonal element $(\mathbb{A}_{\theta})_{pp}$ with the $p$th single particle transition frequency of the quantum NP due to the $\theta$th BEM eigenmode ($\overline{\omega}_p+\lambda_\theta\frac{A_p}{4\pi\overline{\omega}_p}$) and corresponding damping rate ($\frac{\gamma_p}{2}$), while $\overline{\overline{\lambda}}_{pp',\theta\theta}$ constitute coupling matrix elements between such transitions. 
Furthermore, upon introducing the quantized surface charge operator $\hat{q}$ as in Refs.. \cite{fregoni2020,guido2020}, we identify the quantum transition charge sitting on the $k$th tessera as
\begin{equation}\label{eq:SI_q_trans}
\bra{0}\hat{q}_k\ket{\theta,p}=\left(S^{-1/2}T\right)_{k \theta}\sqrt{\frac{A_p}{2\overline{\omega}_p}\left(1+\frac{\lambda_\theta}{2\pi}\right)}
\end{equation}
for each $\ket{\theta,p}$ quantum plasmon mode originating from each $p$th pole of the generic dielectric function for a given $\theta$th BEM eigenmode. Note that simulations reported in main text were performed restricting $\theta$ to the dipolar mode only $\theta_{dip}$, as it is the only relevant mode contributing to the NP optical response in the quasi-static limit. Inclusion of additional modes of higher order is straightforward since each $\theta$th eigenmode leads to an independent response equation which does not couple to the response of other modes (Eq.\ref{eq:SI_mat_cl}). The same approximation has been coherently applied to the classical PCM-NP equations by setting to zero the contribution to the classical response charges from higher-order modes.

To complete the discussion on the mapping between classical and quantum modes, it is useful to consider the perturbation element $\mathbb{V}_{\theta p}$ and how it can be recast following its definition Eq.\ref{eq:SI_gen7_s} and the expression of $\bra{0}\hat{q}_k\ket{\theta,p}$:
\begin{equation}\label{eq:SI_V_trans}
\mathbb{V}_{\theta,p}=\sum_{j}\text{sgn}(A_p)\left(\sqrt{\frac{A_p}{2\overline{\omega}_p}\left(1+\frac{\lambda_\theta}{2\pi}\right)}\right)^*\left(T^{\dagger}S^{-1/2}\right)_{\theta j}V_j=\sum_{j}\text{sgn}(A_p)\bra{\theta,p}\hat{q}_j\ket{0}V_j
\end{equation}
which is the expected form of the perturbation over the tesserae (j index), once the change of sign for negative $A_p$ is accounted for.

On the basis of standard response theory\cite{norman2018,mcweeny}, the response matrix can be conveniently recast in its spectral representation. The generalized eigenvalue problem also discussed in the main text
\begin{equation}\label{eq:SI_response}
\begin{pmatrix}
\mathbb{A_\theta} & \mathbb{B_\theta} \\
\mathbb{B}_\theta^* & \mathbb{A}_\theta^*
\end{pmatrix}
\bm{U}_{\theta}
=\begin{pmatrix}
\mathbb{I} & 0 \\
0 & \mathbb{-I}
\end{pmatrix}
\bm{U}_{\theta}\bm{d}_{\theta}
\end{equation}
specifically reads
\begin{equation}\label{eq:SI_response2}
\left[
\begin{pmatrix}
\mathbb{A_\theta} & \mathbb{B_\theta} \\
\mathbb{B}_\theta^* & \mathbb{A}_\theta^*
\end{pmatrix}
-(\omega_{\theta,n}-i\frac{\gamma_{\theta,n}}{2})
\begin{pmatrix}
\mathbb{I} & 0 \\
0 & \mathbb{-I}
\end{pmatrix}
\right]
\bm{U}_{\theta,n}
=0
\end{equation}
Because of the structure of the response matrix, the eigenvalues come in pairs\cite{mcweeny} (if $\omega_{\theta,n}-i\frac{\gamma_{\theta,n}}{2}$ is an eigenvalue, so it is $-\omega_{\theta,n}-i\frac{\gamma_{\theta,n}}{2}$). Thus it is convenient to use an index $n$ running from 1 to $N$ and from -1 to $-N$ to label them and the eigenvectors: $d_{\theta,n}=\omega_{\theta,n}-i\frac{\gamma_{\theta,n}}{2}$ and $d_{\theta,-n}=-\omega_{\theta,n}-i\frac{\gamma_{\theta,n}}{2}$. The $n-th$ eigenvector, i.e. the $n-th$ column of the matrix  $\bm{U}_{\theta}$, $\bm{U}_{\theta,n}$, has the structure
\begin{equation}\label{eq:SI_response_vectors}
\bm{U}_{\theta,n}=
\begin{pmatrix}
\bm{X}_{\theta,n}\\
\bm{Y}_{\theta,n}
\end{pmatrix}
\; \; \;
\bm{U}_{\theta,-n}=
\begin{pmatrix}
\bm{Y}^*_{\theta,n}\\
\bm{X}^*_{\theta,n}
\end{pmatrix}
\end{equation}
which leads to 
\begin{equation}\label{eq:SI_q_trans_n}
\begin{split}
\bra{0}\hat{q}_k\ket{\theta,n}&=\sum_{p}\left(\bra{0}\hat{q}_k\ket{\theta,p}X_{\theta,pn}+\bra{\theta,p}\hat{q}_k\ket{0}Y_{\theta,pn}\right)= \\
& = \sum_{p}\left(S^{-1/2}T\right)_{k \theta}\left[\sqrt{\frac{A_p}{2\overline{\omega}_p}\left(1+\frac{\lambda_\theta}{2\pi}\right)}X_{\theta,pn} +\left(\sqrt{\frac{A_p}{2\overline{\omega}_p}\left(1+\frac{\lambda_\theta}{2\pi}\right)}\right)^*Y_{\theta,pn} \right] .
\end{split}
\end{equation}
Since Eq.\ref{eq:SI_response} is a generalized eigenvalue problem, its numerical solution is practically obtained by recasting it into a normal eigenvalue problem where the response matrix features a sign change in the lower row
\begin{equation}\label{eq:SI_response3}
\bm{U}^{-1}_{\theta}
\begin{pmatrix}
\mathbb{I} & 0 \\
0 & \mathbb{-I}
\end{pmatrix}
\begin{pmatrix}
\mathbb{A_\theta} & \mathbb{B_\theta} \\
\mathbb{B}_\theta^* & \mathbb{A}_\theta^*
\end{pmatrix}
\bm{U}_{\theta}=\bm{U}^{-1}_{\theta}
\begin{pmatrix}
\mathbb{A_\theta} & \mathbb{B_\theta} \\
-\mathbb{B}_\theta^* & -\mathbb{A}_\theta^*
\end{pmatrix}
\bm{U}_{\theta}=\bm{d}_{\theta}
\end{equation}
which is the actual matrix diagonalization performed in practice. By solving Eq.\ref{eq:SI_mat_cl} for $\mathbb{q}_\theta^{R}=\mathbb{X}_\theta$ and $\mathbb{q}_\theta^{A}=\mathbb{Y}_\theta$, while making use of Eq.\ref{eq:SI_response3} it follows,
\begin{equation}\label{eq:SI_mat_cl_inv}
\begin{pmatrix}
\mathbb{X}_\theta \\
\mathbb{Y}_\theta
\end{pmatrix}
=\bm{U}_{\theta}\left[\bm{d}_{\theta}-\omega
\begin{pmatrix}
\mathbb{I} & 0 \\
0 & \mathbb{I}
\end{pmatrix}
\right]^{-1}
\bm{U}_{\theta}^{-1}
\begin{pmatrix}
\mathbb{I} & 0 \\
0 & \mathbb{-I}
\end{pmatrix}
\begin{pmatrix}
\mathbb{V}_\theta \\
\mathbb{V}_\theta^*
\end{pmatrix}
\end{equation}
Note that due to the non-hermicity of $\mathbb{A}_{\theta}$, $\bm{U}^{-1}_{\theta} \neq \bm{U}_{\theta}^{\dagger}$ and so its $n-th$ row reads
\begin{equation}\label{eq:SI_response_vectors2}
\bm{U}_{\theta,n}^{-1}=
\begin{pmatrix}
\bm{Z}_{\theta,n}\ \bm{W}_{\theta,n}
\end{pmatrix}
\; \; \;
\bm{U}_{\theta,-n}^{-1}=
\begin{pmatrix}
\bm{W}^*_{\theta,n}\ \bm{Z}^*_{\theta,n}
\end{pmatrix}
\end{equation}
allowing the following identification:
\begin{equation}\label{eq:SI_q_trans_n2}
\begin{split}
\bra{\theta,n}\hat{q}_k\ket{0}&=\sum_{p}\left(\bra{\theta,p}\hat{q}_k\ket{0}Z_{\theta,np}-\bra{0}\hat{q}_k\ket{\theta,p}W_{\theta,np}\right)= \\
&= \sum_{p}\left(T^{\dagger}S^{-1/2}\right)_{\theta k}  \left[  \left(\sqrt{\frac{A_p}{2\overline{\omega}_p}\left(1+\frac{\lambda_\theta}{2\pi}\right)}\right)^* Z_{\theta,np} - \sqrt{\frac{A_p}{2\overline{\omega}_p}\left(1+\frac{\lambda_\theta}{2\pi}\right)} W_{\theta,np} \right].
\end{split}
\end{equation}
\newline
Note that in this most general case, rigorously speaking $\bra{0}\hat{q}_k\ket{\theta,n}^*\neq \bra{\theta,n}\hat{q}_k\ket{0}$ (although in practice it holds $\bra{0}\hat{q}_k\ket{\theta,n}^*\approx \bra{\theta,n}\hat{q}_k\ket{0}$); thus the notation should be understood to refer only approximately to a set of excited states $\ket{\theta,n}$. 
\newline On the basis of eqs.\ref{eq:SI_response}-\ref{eq:SI_q_trans_n2} and  response theory\cite{mcweeny}, the spectral representation of the dipole-dipole polarizability of the quantum NP becomes
\begin{eqnarray}\label{eq:SI_pol_gen}
 \nonumber \alpha_{ab}(\omega)=\sum_{\theta,n} \frac{\bra{0}\hat{\mu}_a\ket{\theta,n} \sum_p \text{sgn}(A_p)(\bra{\theta,p}\hat{\mu}_b\ket{0}Z_{\theta,np}-\bra{0}\hat{\mu}_b\ket{\theta,p}W_{\theta,np}  ) }{\omega_{\theta, n}-\omega-i\frac{\gamma_{\theta,n}}{2}}+ \\
 \frac{\bra{0}\hat{\mu}_a\ket{\theta,n}^*\left(\sum_p \text{sgn}(A_p)(\bra{\theta,p}\hat{\mu}_b\ket{0}Z_{\theta,np}-\bra{0}\hat{\mu}_b\ket{\theta,p}W_{\theta,np}  )\right)^*}{\omega_{\theta,n}+\omega+i\frac{\gamma_{\theta,n}}{2}}
\end{eqnarray}
When there are no negative $A_p$, this equation can be further simplify to:
\begin{equation}\label{eq:SI_pol}
 \alpha_{ab}(\omega)=\sum_{\theta,n} \frac{\bra{0}\hat{\mu}_a\ket{\theta,n}\bra{\theta,n}\hat{\mu}_b\ket{0}}{\omega_{\theta, n}-\omega-i\frac{\gamma_{\theta,n}}{2}}+ \\
 \frac{\bra{0}\hat{\mu}_a\ket{\theta,n}^*\bra{\theta,n}\hat{\mu}_b\ket{0}^*}{\omega_{\theta,n}+\omega+i\frac{\gamma_{\theta,n}}{2}}
\end{equation}
where $\hat{\mu}_a=\hat{q}_k\vec{r}_{k,a}$ with $\vec{r}_{k,a}$ being the $a$th component of the position vector pointing to the $k$th surface NP tessera.

\section{Plasmon-molecule Hamiltonian in the generic dielectric function framework}

The Q-PCM-NP quantum model described in SI Section\,\ref{sec:SI_theory} can be coupled to a quantum chemistry molecular description to investigate plexcitonic effects at full-quantum level.

Following Ref.\citenum{fregoni2021} and considering the results of SI\,\ref{sec:SI_theory}, the full plasmon-molecule Hamiltonian in the generic dielectric function Q-PCM-NP approach can be written as
\begin{equation}\label{eq:SI_H_full}
\hat{H}_{\text{tot}}=\hat{H}_{\text{mol}}+\sum_{\theta,n}\left(\omega_{\theta, n}-i\frac{\gamma_{\theta,n}}{2}\right)\hat{b}^{\dagger}_{\theta,n}\hat{b}_{\theta,n}+\sum_{\theta,n,k}\hat{V}_k\left( \bra{\theta,n}\hat{q}_k\ket{0}\hat{b}^{\dagger}_{\theta,n}+  \bra{0}\hat{q}_k\ket{\theta,n}\hat{b}_{\theta,n}\right)
\end{equation}
where $\hat{H}_{\text{mol}}$ is the standard molecular electronic Hamiltonian and $\hat{V}_k$ is the molecular electrostatic potential operator evaluated at centroid of the $k\text{-th}$ surface tessera of the discretized NP.

Solving the full Hamiltonian in Eq.\ref{eq:SI_H_full} at ab initio level is often computationally demanding. When plasmon–molecule interactions and correlation effects are not too strong, a practical approximation is to first solve the molecular Hamiltonian $\hat{H}_{\text{mol}}$ and subsequently couple its eigenstates to the plasmonic system.
Following this line, upon denoting the eigenstates of $\hat{H}_{\text{mol}}$ as $\ket{g}$ for the molecular ground state and $\ket{e}$ for one generic excited state,  $\hat{H}_{\text{tot}}$ can be expressed as,
\begin{eqnarray}\label{eq:SI_H_full_matrix}
\hat{H}_{\text{tot}}=&&\omega_g\ket{g}\bra{g}+\sum_e\left(\omega_e-i\frac{\gamma_e}{2}\right)\ket{e}\bra{e}+\sum_{\theta,n}\left(\omega_{\theta, n}-i\frac{\gamma_{\theta,n}}{2}\right)\hat{b}^{\dagger}_{\theta,n}\hat{b}_{\theta,n}+ \nonumber \\
&& +\sum_{\theta,n,k}\hat{V}_k\left( \bra{\theta,n}\hat{q}_k\ket{0}\hat{b}^{\dagger}_{\theta,n}+  \bra{0}\hat{q}_k\ket{\theta,n}\hat{b}_{\theta,n}\right)
\end{eqnarray}
where $\omega_g$ and $\omega_e$ are the molecular energies of ground and excited states, respectively, while $\gamma_e$ is a phenomenological damping rate to account for the finite lifetime of the molecular excited state $\ket{e}$.

Upon introducing the molecular identity operator $\mathbb{1_{mol}}=\ket{g}\bra{g}+\sum_e\ket{e}\bra{e}$ before and after $\hat{V}_k$ in Eq.\ref{eq:SI_H_full_matrix}, it follows
\begin{eqnarray}\label{eq:SI_H_full_matrix2}
\hat{H}_{\text{tot}}=&&\omega_g\ket{g}\bra{g}+\sum_e\left(\omega_e-i\frac{\gamma_e}{2}\right)\ket{e}\bra{e}+\sum_{\theta,n}\left(\omega_{\theta, n}-i\frac{\gamma_{\theta,n}}{2}\right)\hat{b}^{\dagger}_{\theta,n}\hat{b}_{\theta,n}+\nonumber \\
&&+\sum_{\theta,n,k}\left(\hat{V}_k^{gg}\ket{g}\bra{g}+\sum_e\hat{V}_k^{ee}\ket{e}\bra{e}\right)\left( \bra{\theta,n}\hat{q}_k\ket{0}\hat{b}^{\dagger}_{\theta,n}+  \bra{0}\hat{q}_k\ket{\theta,n}\hat{b}_{\theta,n}\right) \nonumber \\
&& +\sum_{\theta,n,k,e}\left(\hat{V}_k^{eg}\ket{e}\bra{g}+\hat{V}_k^{ge}\ket{g}\bra{e}\right)\left( \bra{\theta,n}\hat{q}_k\ket{0}\hat{b}^{\dagger}_{\theta,n}+  \bra{0}\hat{q}_k\ket{\theta,n}\hat{b}_{\theta,n}\right)
\nonumber \\
&& +
\sum_{\theta,n,k,e \neq e'}\hat{V}_k^{ee'}\ket{e}\bra{e'}\left( \bra{\theta,n}\hat{q}_k\ket{0}\hat{b}^{\dagger}_{\theta,n}+  \bra{0}\hat{q}_k\ket{\theta,n}\hat{b}_{\theta,n}\right)
\end{eqnarray}
where $\hat{V}_k^{ge}$ is a shorthand notation for $\bra{g}\hat{V}_k\ket{e}$. Note that the diagonal terms of the plasmon-molecule interaction (second line of Eq.\ref{eq:SI_H_full_matrix2}) are NP polarization terms that lead to a correction of the bare molecular excitation energies $\omega_e-\omega_g$. The last line of the equation contains terms that promote a mixing of the bare molecular excited states due to molecule-NP polarization effects.

Eq.\ref{eq:SI_H_full_matrix2} can be further simplified upon applying the usual rotating-wave approximation to the  interaction terms in the last two lines of Eq.\ref{eq:SI_H_full_matrix2}, leading to
\begin{eqnarray}\label{eq:SI_H_full_matrix3}
\hat{H}_{\text{tot}}=&&\omega_g\ket{g}\bra{g}+\sum_e\left(\omega_e-i\frac{\gamma_e}{2}\right)\ket{e}\bra{e}+\sum_{\theta,n}\left(\omega_{\theta, n}-i\frac{\gamma_{\theta,n}}{2}\right)\hat{b}^{\dagger}_{\theta,n}\hat{b}_{\theta,n}+\nonumber \\
&&+\sum_{\theta,n,k}\left(\hat{V}_k^{gg}\ket{g}\bra{g}+\sum_e\hat{V}_k^{ee}\ket{e}\bra{e}\right)\left( \bra{\theta,n}\hat{q}_k\ket{0}\hat{b}^{\dagger}_{\theta,n}+  \bra{0}\hat{q}_k\ket{\theta,n}\hat{b}_{\theta,n}\right) \nonumber \\
&& +\sum_{\theta,n,k,e}\hat{V}_{k}^{eg}\left( \bra{\theta,n}\hat{q}_k\ket{0}\hat{b}^{\dagger}_{\theta,n}\ket{g}\bra{e}+\bra{0}\hat{q}_k\ket{\theta,n}\hat{b}_{\theta,n}\ket{e}\bra{g}\right) \nonumber \\
&& +\sum_{\theta,n,k,e<e'}\hat{V}_{k}^{ee'}\left( \bra{\theta,n}\hat{q}_k\ket{0}\hat{b}^{\dagger}_{\theta,n}\ket{e}\bra{e'}+\bra{0}\hat{q}_k\ket{\theta,n}\hat{b}_{\theta,n}\ket{e'}\bra{e}\right) 
\, .
\end{eqnarray}
Note that the molecular transition potential terms are often real in standard quantum chemistry codes and so it holds $\hat{V}_k^{eg}=\hat{V}_k^{ge}$ and $\hat{V}_k^{ee'}=\hat{V}_k^{e'e}$, which is considered in Eq.\ref{eq:SI_H_full_matrix3}. 

In the present work the Hamiltonian of Eq.\ref{eq:SI_H_full_matrix3} is considered and diagonalized in the single-excitation subspace ($\ket{g}\otimes\ket{0_{\theta,n}},\ket{e}\otimes\ket{0_{\theta,n}},\ket{g}\otimes\ket{1_{\theta,n}}$) to investigate plasmon-molecule coupling effects (see main text, Figs.3-5). In such subspace, the interaction terms proportional to $\propto\hat{V}_k^{ee’}$  are dropped because they only couple states of the form ($\ket{e}\otimes\ket{0_{\theta,n}},\ket{e’}\otimes\ket{1_{\theta,n}}$) where the latter term would require extending the subspace to double-excitations. Such approximation is well-justified because such terms are numerically negligible in the system investigated in the present work.

The Hamiltonian of eq.\ref{eq:SI_H_full_matrix3} is the same Hamiltonian discussed in main text (Eq.30).

\section{Computational details}
\begin{figure*}
\centering
\includegraphics[width=1.0\textwidth]{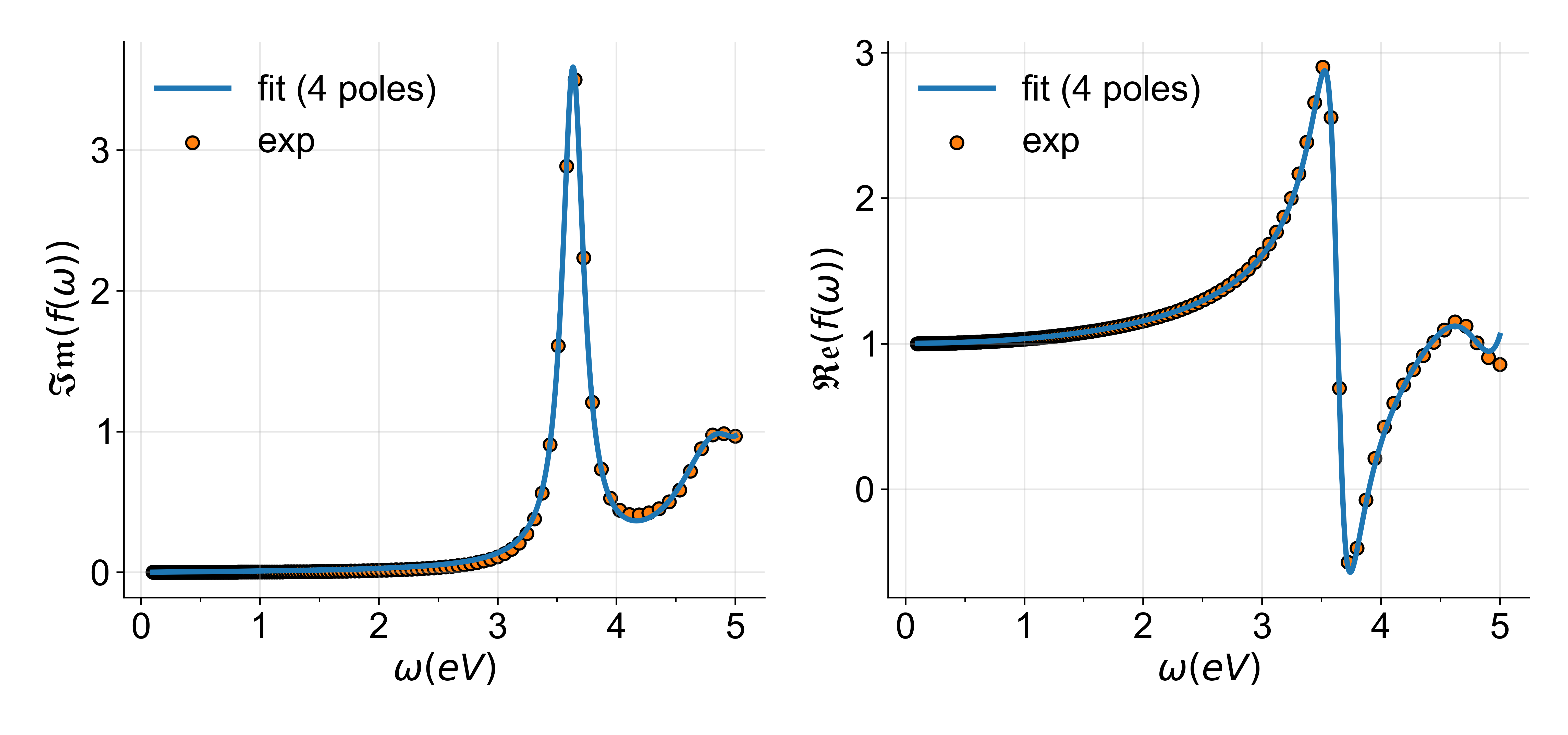}
\caption{ Fitting of Ag $f(\omega)$ function with 4 poles to Brendel-Bormann\cite{rakic1998} reference data (orange dots). The fitting result is the solid blue line. Left and right panels respectively show imaginary and real parts of $f(\omega)$. The fitting parameters are given in Table \ref{table:1}.}
\label{fig:Ag_fit}
\end{figure*}

\begin{figure*}
\centering
\includegraphics[width=1.0\textwidth]{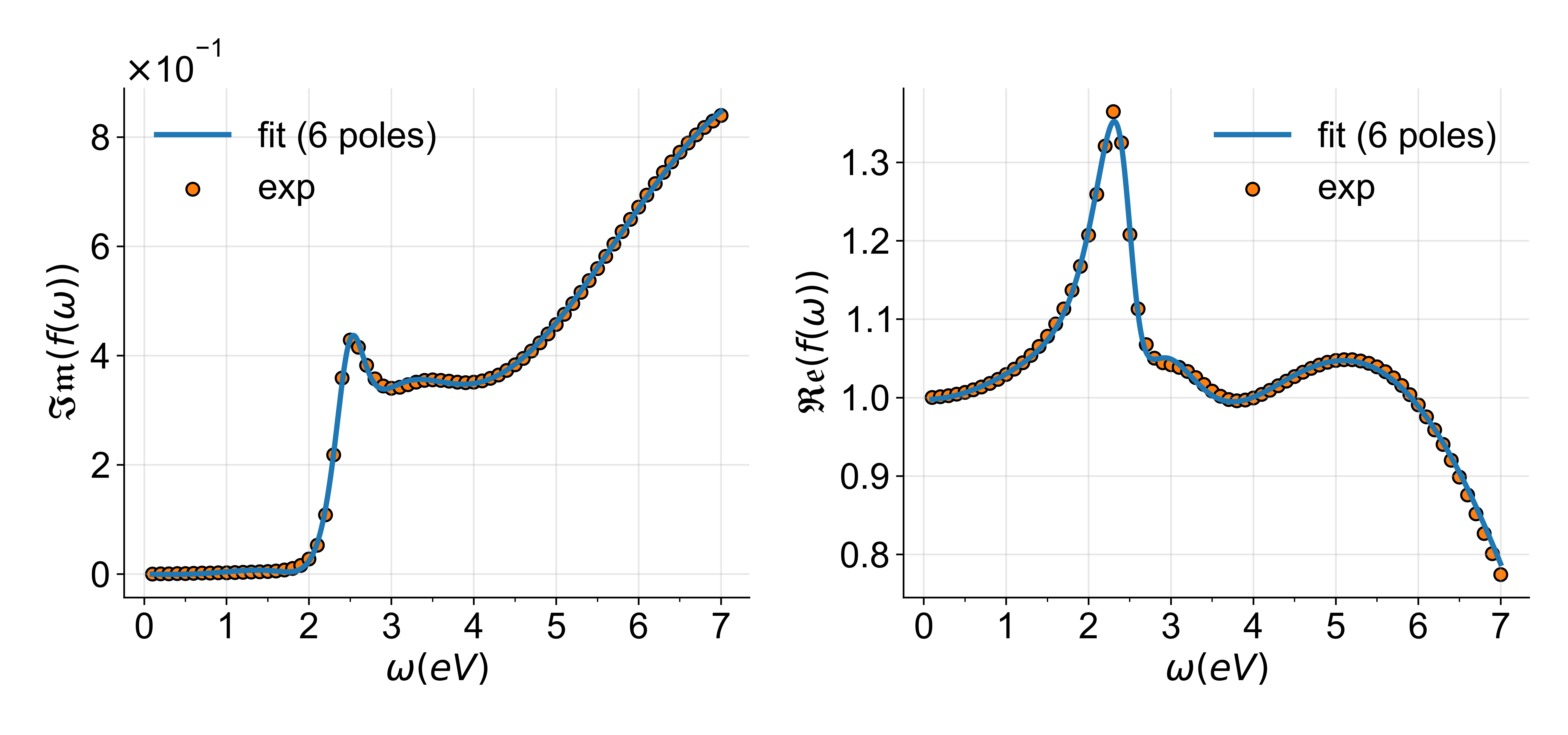}
\caption{Fitting of Au $f(\omega)$ function with 6 poles to Etchegoin\cite{etchegoin2006} reference data (orange dots). The fitting result is the solid blue line. Left and right panels respectively show imaginary and real parts of $f(\omega)$. The fitting parameters are given in Table \ref{table:2}}
\label{fig:Au_fit}
\end{figure*}

Simulations reported in main text have been performed on a test case NP of ellipsoidal shape, whose main axis is $10\,\textrm{nm}$ long and the short axis is $6\,\textrm{nm}$ long (Fig.\,1). 
The surface mesh has been created with the Gmsh\cite{geuzaine2009} code and features 1371 surface tesserae.
The Ag Brendel-Bormann\cite{rakic1998} and Au Etchegoin\cite{etchegoin2006} frequency-dependent data of the metal dielectric functions were fitted as a sum of 4 and 6 DL poles\cite{dall2020real}, respectively.
The fitting $f(\omega)$ functions are shown in Figs. \ref{fig:Ag_fit}-\ref{fig:Au_fit} and the corresponding fitting parameters are given in Tables \ref{table:1}-\ref{table:2}. 
As mentioned in SI 1 only the dipolar mode $\theta=\theta_{\text{dip}}$ has been considered throughout the work since it is the only relevant one for the optical response in the quasi-static limit.

When coupling with the quinolone molecule is investigated, the same ellipsoidal NP shape has been used but the number of tesserae in the surface region closer to the molecule was increased to improve numerically the evaluation of the  the molecule-NP electrostatic coupling terms. 
In that case 2122 tesserae were used. 
The quinolone molecule of Fig.3 (main text) was described at CIS/6-31g** level of theory using a locally modified version of the GAMESS code\cite{schmidt1993general} to obtain the molecular transition potential matrix elements evaluated at the NP surface.

The DL dielectric function that is used for the results of main text Fig.4 is of the form\cite{fregoni2020},
\begin{equation}\label{eq:SI_DL}
\epsilon_{\text{DL}}(\omega)=1+\frac{\Omega_{\text{DL}}^2}{\omega_{\text{DL}}^2-\omega^2-i\gamma_{\text{DL}}\omega}
\end{equation}
and the following parameters $\Omega_{\text{DL}}=4.78\,\text{eV}\,;\omega_{\text{DL}}=2.17\,\text{eV}\,;\gamma_{\text{DL}}=0.219\,\text{eV}$ where used to obtain the Q-PCM-NP DL modes according to the methodology described in Ref.\citenum{fregoni2020}.

The molecular damping rate considered for the results in main text is $\gamma_e=6.6\,\text{meV}\,(\approx 100\text{fs})$.

\begin{table}
\centering
\begin{tabular}{ |c|c|c|c| } 
 \hline
 pole n. & $\omega_p$ (eV) & $\gamma_p$ (eV) & $A_p$ $(\textrm{eV}^2)$ \\
\hline
1 & 3.64 & 0.219 & 2.79 \\
2 & 4.85 & 0.763 & 3.33 \\
3 & 5.18 & 0.0861 & 1.51 \\
4 & 9.89 & 0.0 & 58.1 \\
\hline
\end{tabular}
\caption{Fitting pole parameters of $f(\omega)$ for Ag Brendel-Bormann\cite{rakic1998} reference data.}
\label{table:1}
\end{table}
\newpage
\begin{table}
\centering
\begin{tabular}{ |c|c|c|c| } 
 \hline
 pole n. & $\omega_p$ (eV) & $\gamma_p$ (eV) & $A_p$ $(\textrm{eV}^2)$ \\
\hline
1 & 2.51 & 0.573 & 0.674 \\
2 & 2.59 & 1.76 & -5.19 \\
3 & 2.85 & 2.34 & 7.90 \\
4 & 5.24 & 9.06 & 15.4 \\
5 & 8.39 & 7.52 & 63.1 \\
6 & 42.4 & 0 & 643 \\
\hline
\end{tabular}
\caption{Fitting pole parameters of $f(\omega)$ for Au Etchegoin\cite{etchegoin2006} reference data.}\label{table:2}
\end{table}
\clearpage
\section{Real part of $\alpha_{xx}(\omega)$ for the ellipsoidal NP of Figs.1-2}
\begin{figure*}
\centering
\includegraphics[width=\textwidth]{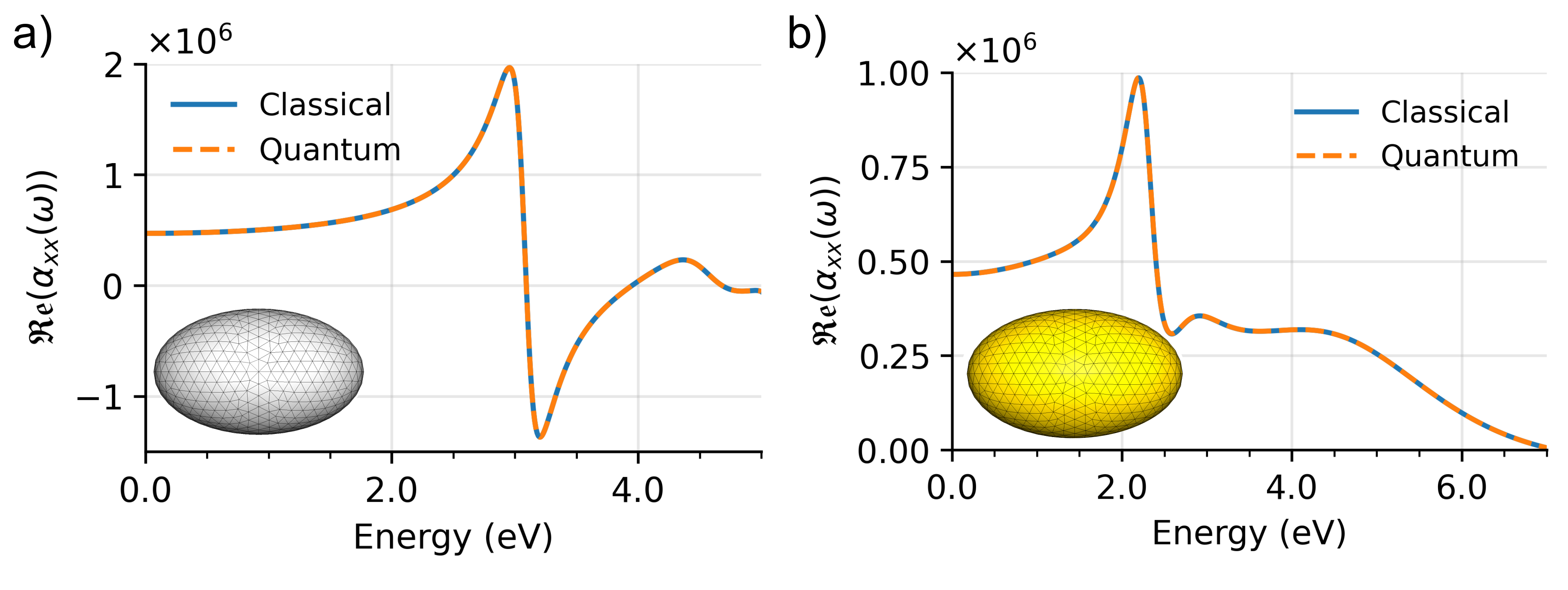}
\caption{Real part of the xx component of the dipole-dipole polarizability tensor for the ellipsoidal NP (Fig.\,1) when the $f(\omega)$ function is fitted to Ag Brendel-Bormann\cite{rakic1998} (a) or Au Etchegoin\cite{etchegoin2006} (b) reference data. The classical PCM-NP (solid blue) and quantum Q-PCM-NP results (dashed orange) are shown.}
\label{fig:AgAu_re}
\end{figure*}
\clearpage
\section{Results for a gold spherical NP}
\begin{figure*}
\centering
\includegraphics[width=\textwidth]{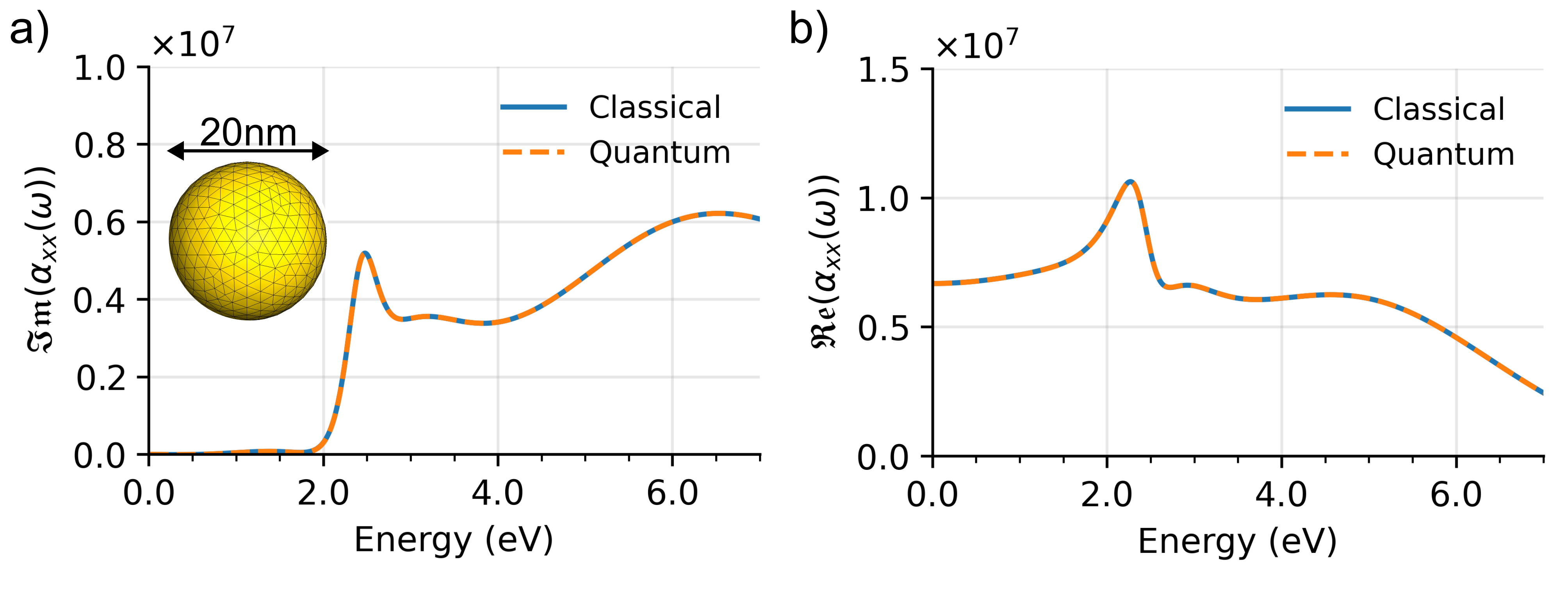}
\caption{1) Imaginary and real part  2) of the xx component of the dipole-dipole polarizability tensor for a gold spherical NP featuring a diameter of 20\,nm NP. The same $f(\omega)$ function as in the case of the ellipsoidal Au NP is used. The classical PCM-NP (solid blue) and quantum Q-PCM-NP results (dashed orange) are shown. Despite spherical NPs have three degenerate dipolar modes only one dipolar mode (oriented along x) is considered here for simplicity. The same holds for the other modes.
}
\label{fig:Au_sphere}
\end{figure*}
\newpage 
\bibliography{bib}